\documentclass[compsoc, conference, a4paper, 10pt, times]{IEEEtran}

\usepackage{amsmath,amsfonts}
\usepackage{algorithmic}
\usepackage{array}
\usepackage{graphicx}
\usepackage{cite}
\usepackage{booktabs}
\usepackage[ruled,vlined]{algorithm2e}
\usepackage{makecell}
\usepackage{array,hhline}
\usepackage{subcaption}
\usepackage{amssymb}
\usepackage{bmpsize}
\usepackage{lipsum}
\usepackage{times}
\usepackage{epsfig}
\usepackage{soul}
\usepackage[utf8]{inputenc}
\usepackage{caption}
\usepackage{enumitem}
\usepackage{adjustbox}
\usepackage{tabu,multirow}
\usepackage{float}
\usepackage{amsthm}
\usepackage{wrapfig,booktabs}

\newtheorem{definition}{Definition}
\newtheorem{theorem}{Theorem}

\usepackage[hidelinks]{hyperref}
\usepackage{xcolor}
\usepackage[misc]{ifsym}

\newenvironment{packeditemize}{
\begin{list}{$\bullet$}{
\setlength{\labelwidth}{8pt}
\setlength{\itemsep}{0pt}
\setlength{\leftmargin}{\labelwidth}
\addtolength{\leftmargin}{\labelsep}
\setlength{\parindent}{0pt}
\setlength{\listparindent}{\parindent}
\setlength{\parsep}{0pt}
\setlength{\topsep}{3pt}}}{\end{list}}

\newcommand{\Name}{$\texttt{CFP-iBDv2}$\xspace}
\newcommand{\Namea}{$\texttt{CFP-AE}$\xspace}
\newcommand{\Nameb}{$\texttt{CFP-iBDv1}$\xspace}

\renewcommand{\figurename}{Fig.}
\renewcommand{\tablename}{Table}

\begin{document}

\title{Fingerprinting Image-to-Image Generative Adversarial Networks}

\author{\IEEEauthorblockN{Guanlin~Li\textsuperscript{1,2}, Guowen~Xu\textsuperscript{3,*,\$}, Han~Qiu\textsuperscript{4}, Shangwei~Guo\textsuperscript{5}, Run~Wang\textsuperscript{6}},
\IEEEauthorblockN{Jiwei Li\textsuperscript{7}, Tianwei~Zhang\textsuperscript{1}, Rongxing~Lu\textsuperscript{8}}
\IEEEauthorblockA{\textit{\textsuperscript{1}Nanyang Technological University, \textsuperscript{2}S-Lab, \textsuperscript{3}City University of Hong Kong, \textsuperscript{4}Tsinghua University}}
\IEEEauthorblockA{\textit{\textsuperscript{5}Chongqing University, \textsuperscript{6}Wuhan University, \textsuperscript{7}Zhejiang University, \textsuperscript{8}University of New Brunswick} \\
guanlin001@e.ntu.edu.sg, guowenxu@cityu.edu.hk, \textsuperscript{*}corresponding author} \textsuperscript{\$}{This work was done when Guowen Xu was working as a research fellow at NTU.}
}

\maketitle

\begin{abstract}

Generative Adversarial Networks (GANs) have been widely used in various application scenarios. Since the production of a commercial GAN requires substantial computational and human resources, the copyright protection of GANs is urgently needed. This paper presents a novel fingerprinting scheme for the Intellectual Property (IP) protection of image-to-image GANs based on a trusted third party. We break through the stealthiness and robustness bottlenecks suffered by previous fingerprinting methods for classification models being naively transferred to GANs. Specifically, we innovatively construct a \textit{composite deep learning model} from the target GAN and a classifier. Then we generate fingerprint samples from this composite model, and embed them in the classifier for effective ownership verification. This scheme inspires some concrete methodologies to practically protect the modern image-to-image translation GANs. Theoretical analysis proves that these methods can satisfy different security requirements necessary for IP protection. We also conduct extensive experiments to show that our solutions outperform existing strategies.
\end{abstract}

\vspace{-5pt}
\section{Introduction}\label{sec:introduction}
\vspace{-5pt}

Generative Adversarial Networks (GANs) for image-to-image (I2I) translation~\cite{stylegan} are used in various applications, e.g., attribute editing~\cite{he_attgan_2019}, domain translation~\cite{cyclegan}, and super resolution~\cite{sr}. 
A well-trained I2I GAN model (especially the generator) is regarded as the core Intellectual Property (IP) due to two reasons~\cite{ong2021protecting}. First, to handle complicated tasks and datasets, modern GAN models are designed to be more sophisticated. For instance, CycleGAN~\cite{cyclegan} and StyleGAN~\cite{stylegan} have 54 and 100 to 300 giga floating-point operations (GFLOPs), depending on hardware and implementations.  
Training such a production-level GAN model usually requires a large amount of computing resources, valuable data, and human expertise. Second, I2I GANs are adopted in many applications with huge commercial values, such as image/video filters in TikTok \cite{TikTok}, Prisma~\cite{Prisma}, and Photoleap~\cite{Photoleap}. So, model vendors have motivations to protect such assets, and prevent malicious model buyers or customers from abusing, copying, \mbox{or redistributing the models without authorization.}

Existing IP protection methods for deep learning models can be roughly divided into two categories. 
(1) \textit{Watermarking}: the model owner embeds carefully-crafted watermarks into the target model by a parameter regularizer~\cite{uchida2017embedding} or backdoor data poisoning~\cite{adi2018turning,le2019adversarial,li2019prove,zhang2018protecting}. Later, the watermarks can be extracted from the model parameters or output as the ownership evidence. 
(2) \textit{Fingerprinting}: the model owner generates unique sample-label pairs that can exactly characterize the target model with a higher probability (Fig.~\ref{fig:finscheme1}). Common approaches~\cite{cao2019ipguard,lukas2019deep,wang2021characteristic,wang2021fingerprinting,peng2022fingerprinting} adopt \textbf{adversarial examples} to identify such fingerprint examples. Compared to watermarking, fingerprinting does not need to modify the target model. Hence, it can better preserve the performance of the target model~\cite{cao2019ipguard,peng2022fingerprinting}. It also shows more applicability and convenience, especially for some scenarios where the model owner does not have the right or capability to modify the models. Due to these advantages, fingerprinting is a more promising method for IP protection of deep {learning models, and we focus on this strategy in this paper.}

However, simply extending prior fingerprinting solutions from classification models to I2I GANs can cause some issues.  
(1) \textit{Persistency}: adversarial examples against GANs are more sensitive to the changes in the model or input-output. So it is easier for an adversary to invalidate such fingerprints by slightly transforming the models or data samples. (2) \textit{Stealthiness}: the adversarial output from a GAN model can be more anomalous than the adversarial label from a classification model, allowing the model thief to detect the fingerprint and then manipulate the verification results. Experiments in Section~\ref{sec:exp} demonstrate these limitations. It is necessary to design a fingerprinting scheme dedicated to I2I GAN models.

We propose the \textit{first} fingerprinting scheme to protect the IP of I2I GAN models \textbf{based on a trusted third party}. The key innovation of our scheme is a \textit{composite deep learning model} constructed from the target GAN model and a classifier (shown in Fig.~\ref{fig:finscheme2}). Specifically, to make the ownership verification \textit{stealthier}, we aim to design a set of fingerprints, whose input samples and output samples from the target model are visually indistinguishable from normal ones. With this requirement, it seems impossible for the model owner to detect the plagiarism, as prior solutions require the output of the plagiarized model has large divergence from the ground truth. To address this issue, we propose to employ a classifier that can accurately identify the output from the plagiarized model, and assign a unique label to it. 
The introduction of the classifier can also enhance the \textit{persistency} of the fingerprints: although the model owner is not permitted to change the target GAN model, he can freely modify the classifier to better recognize the fingerprint output. This benefit cannot be achieved in prior solutions~\cite{cao2019ipguard,lukas2019deep,wang2021characteristic,wang2021fingerprinting}.

Based on this scheme, we provide three concrete designs that can practically protect the IP of GAN models. In the first method (\Namea), the model owner can produce a set of fingerprint samples (i.e., verification samples), whose outputs from the target model are adversarial examples for the owner's classifier, making it give specific labels with a higher probability. In the second and third methods (\Nameb, \Name), the target model's responses to the fingerprint samples are designed to be invisible backdoor samples~\cite{li2019invisible}, which can activate the backdoor embedded in the classifier to produce unique labels. We leverage the Triplet Loss~\cite{schroff_facenet_2015} and fine-grained categorization~\cite{deng2013fine,gavves2013fine} techniques to design novel loss functions, which can implant the backdoor into the classifier for better security and efficiency. 

We perform comprehensive assessments to evaluate our fingerprinting scheme. Specifically, drawing on the core idea of the previous cryptography-based watermarking framework for classification models \cite{adi2018turning}, we theoretically prove that our scheme satisfies four important security requirements: \emph{functionality-preserving}, \emph{unremovability} and \emph{non-rewriteability}. 
Furthermore, through extensive evaluations across three primary I2I tasks (attribute editing, domain translation, super resolution) utilizing advanced GAN models (e.g., AttGAN~\cite{he_attgan_2019}, StarGAN~\cite{choi_stargan_2018}, STGAN~\cite{liu_stgan_2019}), our method demonstrates high versatility and comprehensiveness. Furthermore, our approach surpasses previous strategies in identifying target GAN models and maintaining superior robustness against diverse model and image transformations.


\vspace{-6pt}
\section{Preliminaries}
\label{sec:motivation}
\vspace{-5pt}

\subsection{DNN Fingerprinting}
\label{sec:bg-fp}
\vspace{-5pt}

Fingerprinting is a promising technique to protect the IP of deep learning models~\cite{cao2019ipguard,lukas2019deep,wang2021characteristic,wang2021fingerprinting}. Different from watermarking~\cite{adi2018turning,le2019adversarial,li2019prove,zhang2018protecting}, the model owner constructs the fingerprint and conducts ownership verification without modifying the target model. This brings much more convenience and applicability. 
Researchers proposed solutions to fingerprint classification models with adversarial attacks~\cite{cao2019ipguard,lukas2019deep,wang2021characteristic,wang2021fingerprinting}. The key insight is to craft adversarial examples for the target model, which assigns unique labels to them. During verification, the model owner uses those samples to query a suspicious model. A matched model will give the desired unique labels as ownership evidence with a higher probability. An unrelated model will more probably predict other labels instead of the desired unique labels.

\begin{figure}[t]
     \centering
     \begin{subfigure}[b]{0.45\linewidth}
         \centering
        \includegraphics[width=\linewidth]{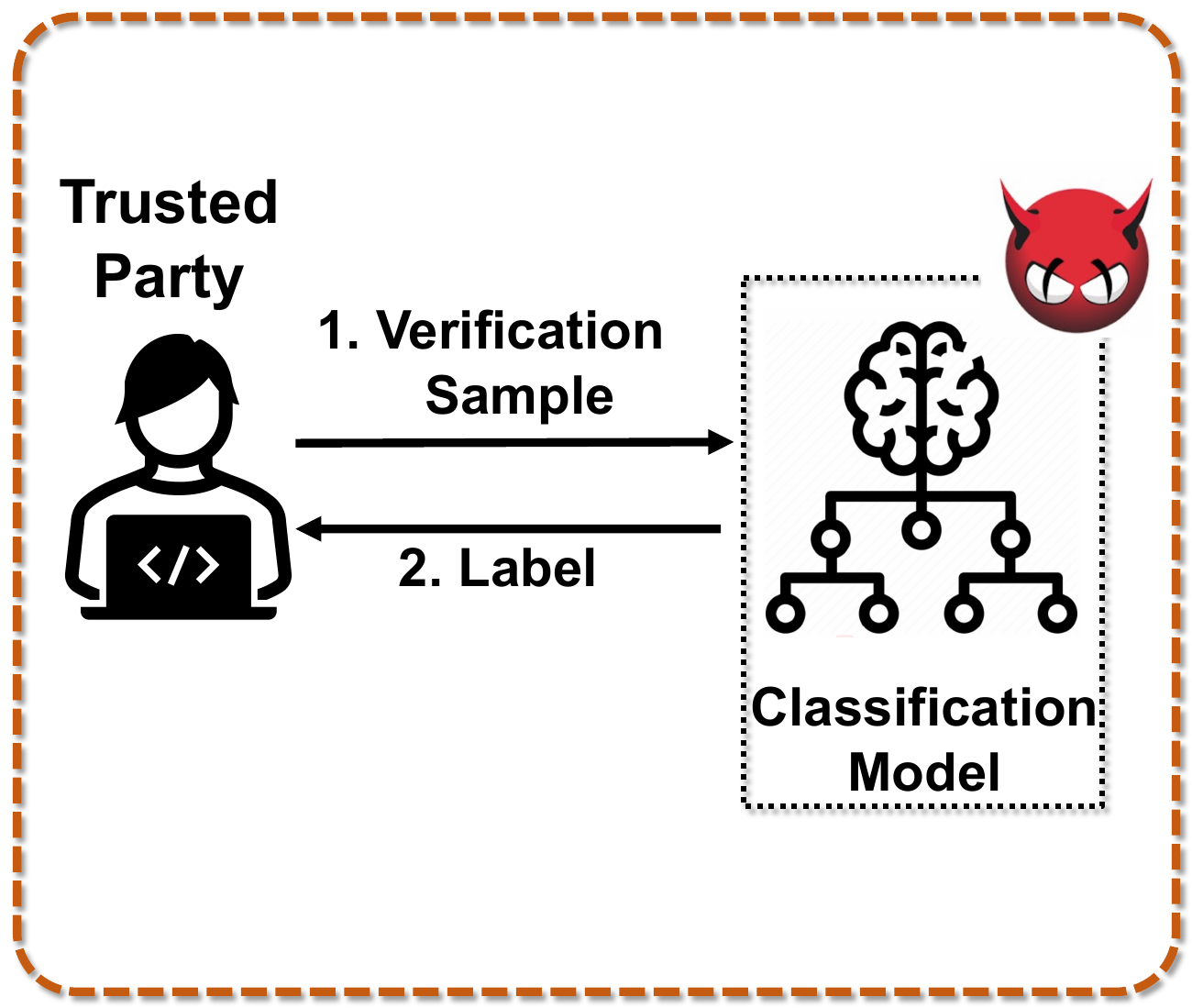} 
         \caption{\mbox{Classification Model}} 
            \label{fig:finscheme1}
     \end{subfigure} \hspace{5pt}
     \begin{subfigure}[b]{0.45\linewidth}
         \centering
        \includegraphics[width=\linewidth]{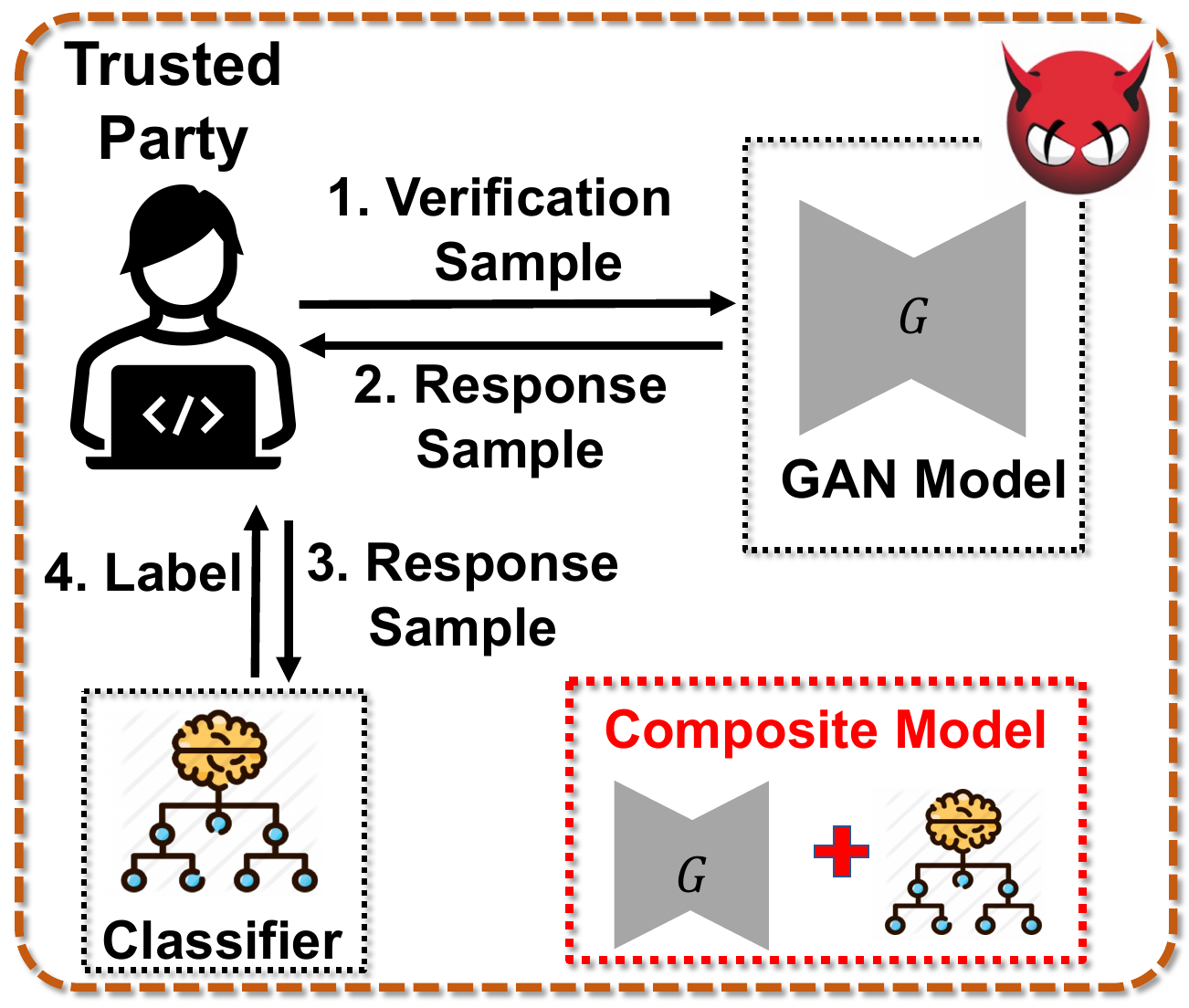}
        \caption{\mbox{GAN Model (ours)}}
         \label{fig:finscheme2}
     \end{subfigure}
    \caption{\textbf{Fingerprinting different kinds of models.}}
        \label{fig:finscheme}
        \vspace{-15pt}
\end{figure}

It becomes difficult when we migrate these strategies to the fingerprinting of I2I GANs. The main difference is that the output of an I2I GAN is images rather than labels. Using adversarial examples of such models for fingerprinting can lead to two problems. First, the fingerprint is less \textit{persistent}: the images generated by GANs are more sensitive to model or input transformations than labels generated by classifiers. An adversary can easily remove the fingerprint from the protected model. Second, the fingerprint is less \textit{stealthy}: a unique label from a classification model is still reasonable, as it belongs to one of possible classes. However, a unique image from a GAN can be suspicious, and easily recognized by the adversary. We will validate these conclusions in Section~\ref{sec:exp}. 

\vspace{-6pt}
\subsection{Commitments}
\label{Commitments}
\vspace{-6pt}

We introduce a trusted third party to help the model owner verify a suspicious model. Considering the potential risk of data leakage and repudiation, it is important to restrict both the trusted third party and users. Therefore, we adopt the commitment scheme \cite{juels1999fuzzy} to implement our verification protocol. It is a widely used cryptographic primitive that allows the sender to lock a secret $x$ in a vault that is free of cryptographic information leakage and tamper-proof, and then send it to others (i.e., a receiver). Generally, a commitment scheme contains two algorithms:

\begin{packeditemize}
\item $ \mathbf{Com}(x, h)$: Given a secret $x\in S$ and a random bit string $h\in \{ 0,1 \}^n$, outputs a bit string $c_x$, where  $h$ transforms $x$ into the ciphertext state.  $S$ represents the value space of x.

\item $ \mathbf{Open}(c_x, x,  h)$: Given a secret $x\in S$, a random bit string $h\in \{ 0,1 \}^n$, and $c_x\in \{0, 1\}^\mathcal{Z}$, if $ \mathbf{Com}(x, h)=c_x$, outputs 1. Otherwise, outputs 0. 
\end{packeditemize}

A commitment scheme has properties: 

\textbf{Correctness:} it is required that for $ {\forall}x\in S$, we have
 \begin{equation*}
\begin{split}
\mathop{Pr}\limits_{h\in \{0, 1\}^n}[\mathbf{Open}(c_x, x,  h)=1|\mathbf{Com}(x, h)\rightarrow c_x]=1
\end{split}
\end{equation*}

\textbf{Binding}: it is impossible for the sender to change the locked secret $x$ once it is sent out.  For any PPT algorithm\footnote{PPT means an algorithm running in probabilistic polynomial time.} $\mathcal{A}$, we have
\begin{equation*}
\mathop{Pr}\left[
\begin{array}{c|c}
 &c_x\leftarrow\mathbf{Com}(x, h)\wedge\\
\mathbf{Open}(c_x, \tilde {x},  \tilde{h})=1 &(\tilde {x},  \tilde{h})\leftarrow \mathcal{A}(c_x, x, h) \wedge\\
&(x, h)\neq  (\tilde {x},  \tilde{h})
\end{array}\right]\leq \epsilon(n),
\end{equation*}
where $\epsilon(n)$ is negligible in $n$ and the probability is taken over $x\in S$ and $h\in \{0,1\}^n$. 

\textbf{Hiding}: it is infeasible for receivers to open the locked $x$ without the sender's help. It requires that no PPT algorithm $\mathcal{A}$ can distinguish $c_{x'}\leftarrow \mathbf{Com}(x', h)$ from $c_x\leftarrow \mathbf{Com}(x, h)$ for any $x, x'\in S$ and $h\in \{ 0,1 \}^n$, where $x\neq x'$.  If for all  $x\neq x'$, the distributions of $c_x$ and $c_{x'}$ are statistically close, i.e.,  $\Delta(c_x, c_{x'})=\frac{1}{2}\sum_{c\in \mathcal{Z}}Pr(c_x=c)-Pr(c_{x'}=c)$  is a negligible function in $n$, where $\mathcal{Z}$ denotes the range space of $c_x$,  we call the commitment scheme statistical hiding. 

\begin{figure*}[h]
\centering
\vspace{-10pt}
\includegraphics[width=1.0\linewidth]{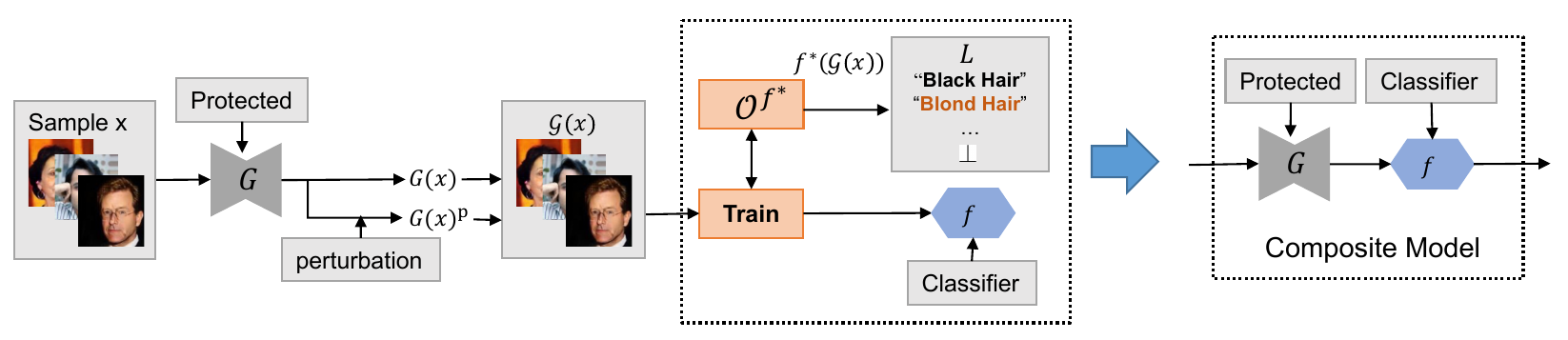}
\caption{\textbf{Training a composite deep learning model.}}
\label{Training process}
\vspace{-10pt}
\end{figure*}

\begin{figure}[h]
\centering
\includegraphics[width=1.0\linewidth]{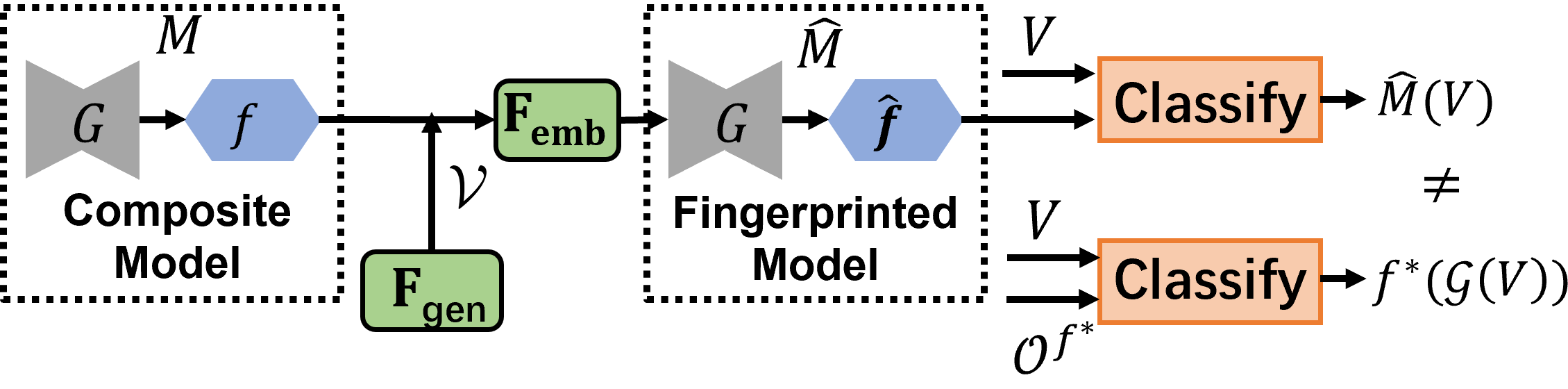}
\caption{\textbf{Generating a fingerprinted composite model.}}
\label{Training process fp}
\vspace{-13pt}
\end{figure}

\vspace{-5pt}
\section{A Novel Fingerprinting Scheme}
\label{sec:preliminary}
\vspace{-5pt}

\subsection{Design Insight}
\vspace{-5pt}

As discussed above, to make the verification stealthier and more indistinguishable from normal inference, \emph{the fingerprint samples and the corresponding model output should be identical to normal cases}. Besides, \textit{the model output should also be unique to differentiate the target and other unrelated models.} Although these two conditions seem to contradict each other, we propose a new scheme to achieve a satisfied tradeoff between them. The general idea is that we craft fingerprint samples with the model output visually similar to normal ones, and employ a classifier to tell if the output is from a target model or not (Fig.~\ref{fig:finscheme2}). A matched GAN model will produce visually normal output samples, but assigned unique labels by the classifier. Below, we describe the detailed steps of our scheme. For the convenience of readers, we show all important symbols used in our paper and their remarks in Appendix~\ref{appendix:sr}.

\vspace{-7pt}
\subsection{Scheme Overview}
\vspace{-7pt}

We consider an I2I GAN model for IP protection. We introduce an additional classifier for ownership judgement, which forms a \textit{composite deep learning model} with the target GAN. Then we carefully craft fingerprints and embed them into the composite model. This process requires that the  embedded fingerprint should be difficult to remove, even if the adversary modifies the GAN model or samples.

We borrow the basic framework from~\cite{adi2018turning}, which is a standard theoretical analysis of DNN watermarks. As we focus on GAN fingerprinting without any model modification, we need to modify this framework to adapt to this requirement. Below, we first give the formal definitions of the composite deep learning model and the fingerprint.  
Based on these, 
we give the workflow of our scheme. For simplicity, we use $n\in\mathbb{N}$ as a security parameter, which is implicit in the input of all algorithms below.   $[k]$ is the shorthand $\{1, 2, \cdots k\}$ for $k\in \mathbb{N}$.

\vspace{-7pt}
\subsection{Composite Deep Learning  Model}
\vspace{-7pt}

We consider a target GAN model $G$ for protection\footnote{Here $G$ is only the generator of the GAN model, as the discriminator is deprecated after the GAN is trained.}, which maps a sample $x\in D$ to another sample $x'\in D$. Here $D \subset \{0,1\}^*$ is the sample space. We introduce a label space $L\subset  \{0,1\}^*\cup\{\perp\}$ for any sample in $D$, which defines the possible properties of the samples generated by $G$, e.g., objects, scenes or conditions in the image. We define $|D|=\Theta (2^n)$ and $|L|=\Omega (p(n))$  for a positive polynomial $p(\cdot)$. A composite deep learning model is defined as below:

\begin{definition}(Composite Deep Learning Model)
Given the GAN model $G$ and its sample space $D$, let $f^*$ be a ground-truth function which classifies a sample $x\in D$ according to its label $y\in L$. Let $\mathcal{G}(x)=\{G(x)\cup G(x)^p| x\in D \}$ be the augmented set of model outputs, where $G(x)$ and $G(x)^p$ denote the accurate model outputs and possible perturbed ones\footnote{$G(x)^p$ is used for training to improve the classification accuracy of $f$ even on the (subtle) perturbations of $G(x)$. Usually, the perturbation could be random noise, random flipping, random cropping, or random rotation, which are widely used in training a classification model. As augmentation methods, we force the classifier to give a correct label to the perturbed images.}. We use the $\mathbf{Train}$ algorithm described below to obtain a classifier $f$, which approximates the mapping: $\mathcal{G}(x)\rightarrow f^*(\mathcal{G}(x))$. Then a composite deep learning model is defined as $M(x) = f(G(x))$.
\end{definition}

\renewcommand{\arraystretch}{1}

The composite model is essentially a mapping $M: D\rightarrow L$, which simulates how humans assign specific labels to GAN-generated samples. To produce the composite model from $G$ and $f^*$, we consider an oracle $\mathcal{O}^{f^*}$, which truly answers each call to $f^*$. Then we have:

\begin{packeditemize}
\item $\mathbf{Train}$($\mathcal{O}^{f^*}$, $\mathcal{G}$): it is a PPT algorithm used to output a classifier $f$, in which $\mathcal{O}^{f^*}$ plays a role like a model training algorithm containing dataset and other necessary components to train a classification model.
\item  $\mathbf{Classify}$($M$, $x$): it is a deterministic function that outputs a \mbox{value $M(x)\in L \backslash \{\perp\}$ for a given input $x\in D$.}

\end{packeditemize}

Fig.~\ref{Training process} gives an example of training a composite deep learning model. We use $\bar{D}=\{x\in D|M(x)\neq \perp\}$ to denote the set of all inputs whose relationship with the output is defined, where $\perp$ stands for out-of-domain cases. Then we say the algorithms ($\mathbf{Train}$, $\mathbf{Classify}$) are $\epsilon$-accurate if $Pr[f^*(\mathcal{G}(x))\neq \mathbf{Classify}(M, x)|x\in \bar{D}]\leq \epsilon$, where the probability arises from the randomness of $\mathbf{Train}$. Thus, we measure accuracy mainly for those inputs that are meaningful to the outputs. For those inputs not defined by the ground-truth classifier $f^*$, we assume their labels are random, i.e.,  for all $x\in D\backslash \bar{D}$ {and any $i\in L$, we have $Pr[\mathbf{Classify}(M, x)=i]=1/|L|$.}

\vspace{-6pt}
\subsection{Fingerprints in Composite Models}
\label{Fingerprints in Composite Deep Learning Model}
\vspace{-6pt}

Our fingerprinting scheme crafts a set of verification samples and a classifier, such that the classifier can assign unique labels to the target model's outputs of these verification samples. Formally, we have:

\begin{definition}(Fingerprint Set for a Composite Model)
\label{fin-cm}
A fingerprint set $\mathcal{V}$ for a composite model $M$ is defined as $(V, V_L)$, where the verification sample set $V\subset D$ and verification label set $V_L\subset $ $L\backslash \{\perp\}$ satisfy the condition: for $x\in V$, $V_L(x)\neq f^{*}(\mathcal{G}(x))$.
\end{definition}

We use an algorithm $\mathbf{F_{gen}}$ to generate a fingerprint set\footnote{Whenever we fix a verification sample set $V$, the fingerprint set implies the corresponding $V_L$.} from the GAN model $G$ and oracle $\mathcal{O}^{f^{*}}$. We further define a PPT algorithm $\mathbf{F_{emb}}$ to embed the generated fingerprint into the composite model. Specifically, given the oracle $\mathcal{O}^{f^{*}}$,  a fingerprint set $\mathcal{V}$, and a composite model $M$, $\mathbf{F_{emb}}$ produces a fingerprinted model $\hat{M}=\hat{f}(G(\cdot))$, which can correctly classify the verification samples $V$ as $V_L$ with a high probability (Fig.~\ref{Training process fp}). Formally, we have: 

\begin{definition}(Fingerprinted Model)
\label{fin-model}
We say a composite model $\hat{M}$ is fingerprinted by $\mathbf{F_{emb}}$, if it behaves like $f^{*}(\mathcal{G}(\cdot))$ on $\bar{D}\backslash V$, and reliably predicts unique labels $V_L$ on $V$, i.e.,
 \begin{equation}
 \label{eq1}
\begin{split}
&\mathop{Pr}\limits_{x\in \bar{D}\backslash V}[f^{*}(\mathcal{G}(x))\neq \mathbf{Classify}(\hat{M}, x)]\leq \epsilon, \mathrm{and} \\
& \mathop{Pr}\limits_{x\in V}[V_L(x)\neq \mathbf{Classify}(\hat{M}, x)]\leq \epsilon.
\end{split}
\end{equation}
\end{definition}

\textit{Remark}:  since a given model may be suspected of being embedded with fingerprints, a strong fingerprint  should be difficult to be reconstructed or be detected by adversaries in arbitrary ways. It requires the fingerprints to satisfy additional requirements to endure various types of attacks. For legibility, we will present these requirements in Section~\ref{Strong Fingerprints}.

\vspace{-6pt}
\subsection{Threat Model}
\label{sec:bg-threat}
\vspace{-6pt}

We exactly follow the \textit{standard} threat model in prior IP protection works \cite{adi2018turning,le2019adversarial,li2019prove,zhang2018protecting,cao2019ipguard,lukas2019deep,wang2021characteristic,wang2021fingerprinting}. It encompasses four distinct identities: model owner, model buyer, adversary, and trusted third party. Specifically, the model owner has invested substantial resources into training a valuable production-level GAN model $G$, using a private internal dataset. The model buyer purchases $G$ from the model owner and adheres to stipulated usage policies, including forbidding model redistribution or resale. The adversary, on the other hand, could be a hacker that attempts to steal $G$ from the buyer, or a dishonest buyer who violates the usage policies, such as illegally reselling the model. The primary objective of the model owner is to discern whether a suspicious model $G^s$ was illegally redistributed based on $G$ or stolen from $G$, employing an advanced fingerprinting scheme. This verification process is assisted by a trusted third party. Basically the model owner registers his model $G$ with the trusted third party by securely sharing the verification samples and classifier $\hat{f}$. With such information, the trusted third party can determine whether a suspicious model is the model owner's property $G$. The verification results could serve as judicial evidence for legitimate purposes.

We make some practical assumptions. First, we assume the suspicious model $G^s$ is deployed as an online service (e.g., \cite{TikTok,MakeGirlsMoe,BeautyCam}), so both the model owner and trusted third party have \textbf{black-box access} to it, i.e., they can only send arbitrary inputs to $G^s$ and receive the corresponding outputs without knowing model parameters and other details. Second, the adversary can alter his model's weights or inference samples, attempting to break the model's fingerprint without decreasing the model's performance. 
Moreover, the adversary can overwrite the fingerprint by launching a new verification process. This new verification process must also be registered with the trusted third party, otherwise, the verification results cannot be recognized as legally effective.

The strong adversarial capability requires the model owner to design a robust fingerprinting scheme against various alterations and evasions. Specifically, we consider three mainstream model transformations (pruning, fine-tuning, and quantizing) and eight image transformations (adding noise, blurring, compressing, cropping, adjusting brightness, adjusting contrast, adjusting gamma, and adjusting hue), that could be potentially used by the adversary. The proposed scheme should be robust against these transformations. On the other hand, the fingerprinting scheme should be visually stealthy and cannot be detected with deep-learning models. To reduce the false alarms on an innocent suspicious model owner, the verification samples should be highly unique to each model $G$.

\begin{figure}[t]
     \centering
         \centering
        \includegraphics[width=\linewidth]{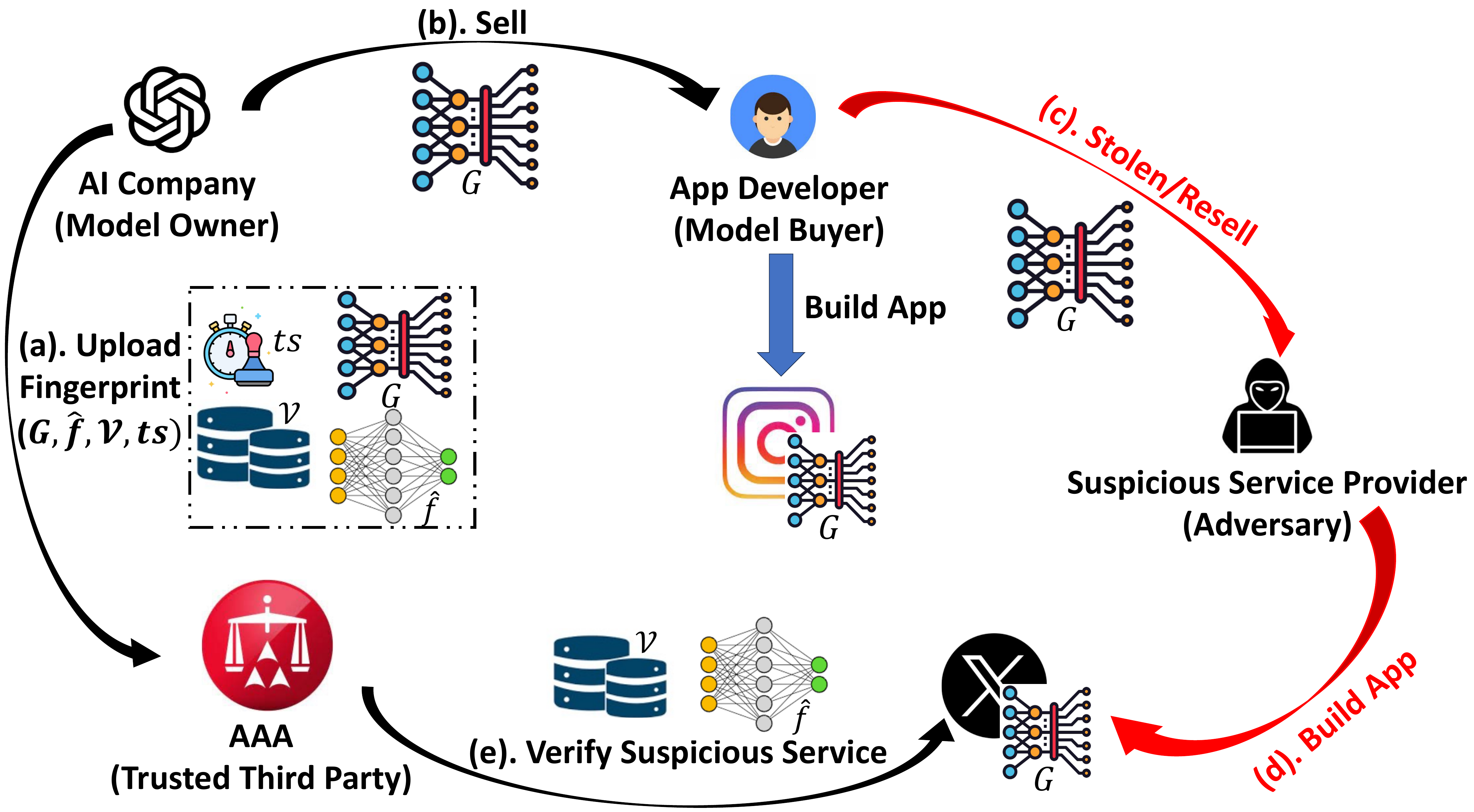} 
    \caption{A real-world scenario.}
        \label{fig:real}
        \vspace{-15pt}
\end{figure}

\vspace{-7pt}
\subsection{A Motivating Example}
\vspace{-7pt}

We provide a motivating example to describe the end-to-end IP protection process, as shown in Fig.~\ref{fig:real}. An AI tech company (e.g., DeepX\footnote{\url{https://deepxhub.com}}, Runwayml\footnote{\url{https://runwayml.com/}}, Saleforce\footnote{\url{https://www.salesforce.com/ap/}}) runs the business of training deep learning models for customers, serving as the \textit{model owner} in our setting. This company trains an I2I GAN $G$, capable of modifying a given image, and sells it to any interested customers. The company wants to protect the IP of its GAN model from any unauthorized redistribution using our fingerprinting scheme. To achieve this, it generates a set of verification samples $\mathcal{V}=(V, V_L)$ and submits all the required components including $G$ and $\hat{f}$ to a \textit{trusted third party}, such as American Arbitration Association (AAA)\footnote{\url{https://www.adr.org/}}, Ohalo\footnote{\url{https://www.ohalo.co/}}, or Dentons Rodyk\footnote{\url{https://dentons.rodyk.com/}}, with a timestamp $ts$ (step \textbf{(a)}). This timestamp is a plaintext followed by a signature to avoid illegal modification. After receiving all the information, the trusted third party launches the verification process on the protected model $G$, verifies the results, and checks whether the plaintext $ts$ is correct. After all the checking, the model $G$ can be safely released for public purchase.

A mobile app developer is developing a photo editing app and is interested in this GAN model $G$. He purchases $G$ from the AI company and integrates it into his app (step \textbf{(b)}). He is thus the \textit{model buyer} in our setting. He needs to follow the usage policy from the model owner that this model cannot be redistributed or resold to other entities. However, an adversary could illegally obtain this model (step \textbf{(c)}), and deploy it in his own online image editing service (step \textbf{(d)}). This could be realized by hacking into the developer's app or the developer dishonestly reselling the model to another party for profit. Being aware of IP protection, the adversary can try different ways to disable the fingerprinting scheme: (1) he can perform different model transformations over $G$; (2) he can perform different image transformations over the input or output images of $G$; (3) he can try to build machine learning models to detect the possible verification samples and then manipulate the output; (4) he can launch a new verification process to overwrite the fingerprint. For any action, he should maintain the normal functionality of the model $G$.


When the AI tech company (model owner) discovers a suspicious image editing service $G^s$ that possibly uses its model $G$ without authorization, it will delegate the trusted third party to execute the verification process on $G^s$ (step \textbf{(e)}). If the verification result suggests that $G^s$ is plagiarized from $G$, it can be used as the judicial evidence for the model owner to sue the provider of $G^s$ (adversary). As mentioned above, the adversary could register the stolen model to the trusted third party to overwrite the fingerprint. However, by verifying and comparing the timestamps of the model owner's registration and adversary's registration, the trusted third party is able to tell if the model owner's fingerprint is overwritten.


\vspace{-7pt}
\subsection{Workflow of Our Fingerprinting Scheme}
\vspace{-7pt}

\begin{figure}[t]
\centering
\includegraphics[width=0.8\linewidth]{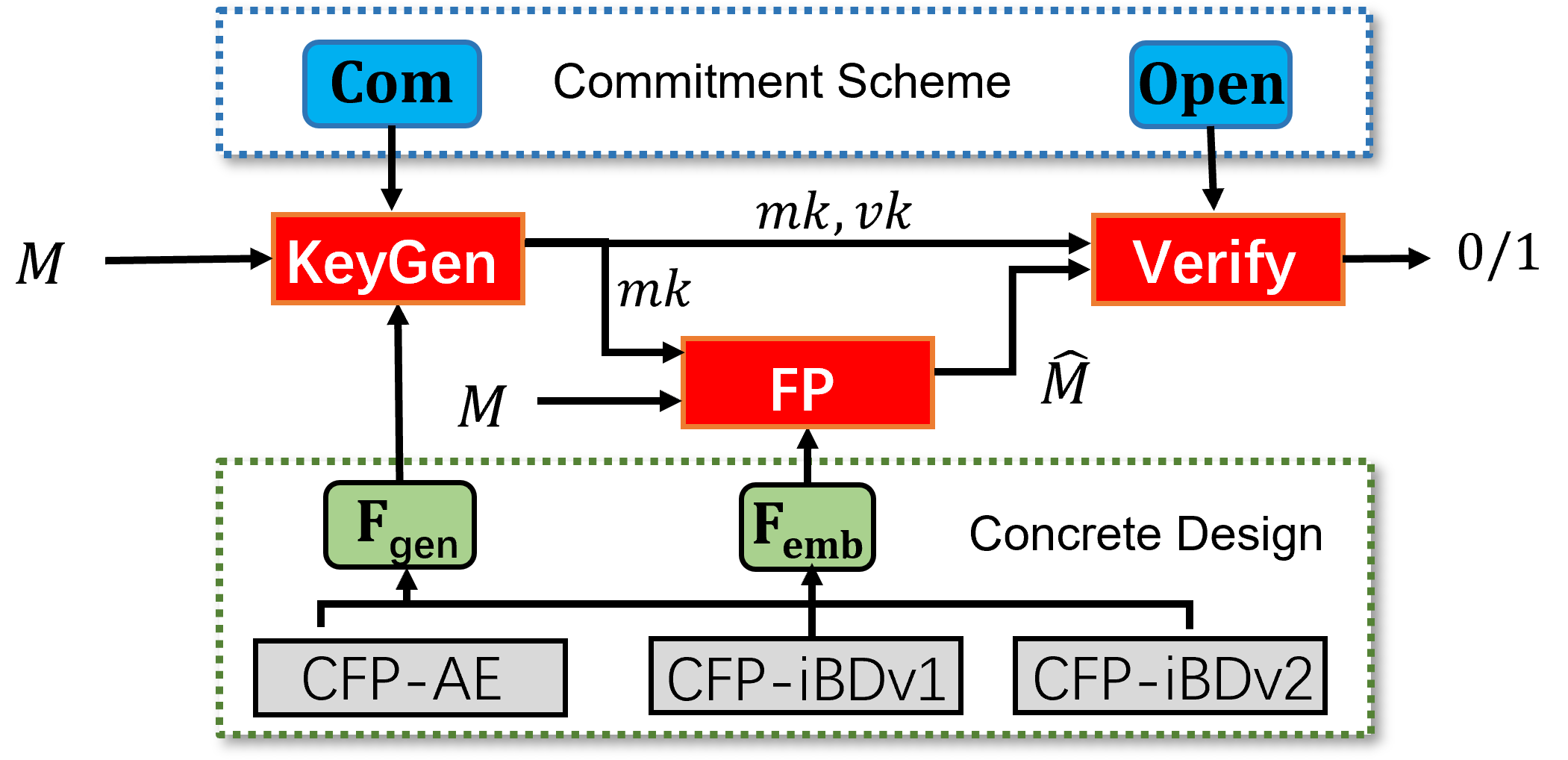}
\caption{\textbf{The workflow of our fingerprinting scheme.}}
\label{The workflow of fingerprint generation and verification}
\vspace{-20pt}
\end{figure}

We now outline our fingerprinting process, as shown in Fig.~\ref{The workflow of fingerprint generation and verification}. Given the targeted model $G$, the model owner first adopts the algorithm $\mathbf{Train}$ to establish the composite deep learning model $M$. Then he uses a series of algorithms to generate a secret marking key $mk$  and a public verification key $vk$, and embed the fingerprint from $mk$ into the model. During verification, the model owner uses marking and verification keys to verify whether a suspicious model contains the fingerprints. Additionally, if the adversary overwrites the fingerprint, we launch a comparison algorithm to resolve the fingerprint conflicts. The entire workflow can be described by four high-level PPT algorithms ($\mathbf{KeyGen}$, $\mathbf{FP}$, $\mathbf{Verify}$, $\mathbf{Compare}$): 
\begin{packeditemize}
\item  $\mathbf{KeyGen}()$: Given a security parameter $n$ and the information related to the model, it outputs the secret marking key $mk$ and the public verification key $vk$, where $mk$ contains the fingerprint to be embedded into the target model, and $vk$ is used for subsequent verification.  This process requires $\mathbf{F_{gen}}$ to generate fingerprint sets. It also requires $\mathbf{Com}$ to commit to the elements in each fingerprint set and random elements selected by the model owner, which provides arguments for subsequent verification.

\item  $\mathbf{FP}$($M, mk$): Given a composite model $M$ and the marking key $mk$, it outputs a fingerprinted model $\hat{M}$.
It uses $\mathbf{F_{emb}}$ as the subroutine to convert $M$ to $\hat{M}$, thereby embedding the fingerprint contained in $mk$ into $M$. Then, a private key $p_k$ and a public key $c_k$ are generated to run a signature algorithm on the current timestamp $ts$ to obtain the signature $s_{ts}$. Finally, $G$, $\hat{f}$, $vk$, $ts||s$, and $c_k$ are sent to the trusted third party.

\item $\mathbf{Verify}$($mk, vk, M$): Given the key pair $mk$, $vk$ and a model $M$, it outputs a bit $b\in\{ 0,1\}$, where 1 means that the verified model has copyright infringement, and vice versa. It uses $\mathbf{Open}$ as the subroutine to open the previous commitments to all the elements in $mk$.

\item $\mathbf{Compare}$($ts||s, c_k, ts'||s', c_k'$): Given two signed timestamps, $ts||s$ and $ts'||s'$, and the corresponding public keys, $c_k$ and $c_k'$, it outputs one bit $\{0,1\}$, where 1 stands for $ts$ earlier, and 0 stands for $ts$ later. It checks the signature for each timestamp with its key.
\end{packeditemize}

Fig.~\ref{Concrete Construction of fingerprinting process} details the algorithms ($\mathbf{KeyGen}$, $\mathbf{FP}$, $\mathbf{Verify}$, $\mathbf{Compare}$) for this process. Specifically, let ($\mathbf{Train}$, $\mathbf{Classify}$) be an $\epsilon$-accurate  composite deep learning  model,  $\mathbf{F_{emb}}$ be a  strong fingerprinting algorithm and ($\mathbf{Com},\mathbf{Open}$) be a statistically hiding commitment scheme. (1) $\mathbf{KeyGen}$ generates strong fingerprints ($\mathbf{F_{gen}}$), which are also used as the secret marking key $mk$. Then a commitment scheme ($\mathbf{Com}$) is used to generate the verification key  $vk$ corresponding to $mk$ for the legitimacy verification of suspicious models.  (2) $\mathbf{FP}$ embeds the fingerprints into the composite model ($\mathbf{F_{emb}}$) and sends all components to the trusted third party with a timestamp $ts$. (3)  $\mathbf{Verify}$ opens the commitments ($\mathbf{Open}$) to all the elements in the secret key $mk$, and uses it to verify whether a suspicious model matches the fingerprints ($\mathbf{Classify}$). If most verification samples in the fingerprint set are predicted as the verification labels by the classifier $\hat{f}$, we infer this GAN model is infringing. When the adversary provides a verification result with the help of the trusted third party as well, $\mathbf{Compare}$ will check the legality of timestamps saved in the trusted third party for the model owner and adversary, and compare their order in the timeline to make a final decision about the ownership of the model. Note that both $\mathbf{Verify}$ and $\mathbf{Compare}$ are completed with the help of a trusted third party.

\vspace{-6pt}
\subsection{Security  Requirements}
\label{security requirements}
\vspace{-6pt}

The correctness of our fingerprinting scheme, i.e., three PPT algorithms ($\mathbf{KeyGen}$, $\mathbf{FP}$, $\mathbf{Verify}$, $\mathbf{Compare}$), requires that for the honestly generated keys $mk, vk$,  and  $ts<ts'$, we have
 \begin{equation*}
\begin{split}
&\mathop{Pr}\limits_{(M, \hat{M}, mk, vk)}[\mathbf{Verify}(mk, vk, \hat{M})=1]=1, and\\
&\mathop{Pr}\limits_{(ts||s, c_k, ts'||s', c_k')}[\mathbf{Compare}(ts||s, c_k, ts'||s', c_k')=1]=1
\end{split}
\end{equation*}
We define following three security requirements: 

(I) \textbf{Functionality-preserving}. This property is twofold in our fingerprinting scheme. The model with fingerprints should be as accurate as the model without fingerprints for classifying normal samples. The proposed scheme should also correctly classify the verification samples $V$ as $V_L$ with a high probability. Formally, for any ($\hat{M}, mk, vk$) honestly generated through the previously described algorithms, it holds that
 \begin{equation*}
 \label{eq1_new}
\begin{split}
&\mathop{Pr}\limits_{x\in \bar{D}\backslash V}[f^{*}(\mathcal{G}(x))\neq \mathbf{Classify}(\hat{M}, x)]\leq \epsilon, \mathrm{and} \\
& \mathop{Pr}\limits_{x\in V}[V_L(x)\neq \mathbf{Classify}(\hat{M}, x)]\leq \epsilon.
\end{split}
\end{equation*}
The first part indicates the fingerprinting scheme does not affect the composite model on non-verification samples with a high probability. The second part indicates the scheme makes the composite model give specific labels to verification samples with a high probability.

(II) \textbf{Unremovability}. This means that the adversary cannot remove the fingerprint even if he knows its existence and the algorithms used. Formally, a fingerprinting scheme with unremovability requires that any PPT algorithm $\mathcal{A}$ wins the following \textbf{Game 1} only with a negligible probability\footnote{To facilitate the security proof, we assume the adversary has access to the full composite model. In practice, the model owner only releases the GAN part of the composite model, i.e., $G$, and retains the classifier part for subsequent verification. This does not affect the proof because if the adversary cannot win the game with full access to the composite model, it is less capable to gain an advantage with weaker prior knowledge.}.
\begin{itemize}
\setlength\itemsep{0.2em}
\item [1.] Generate $M\leftarrow \mathbf{Train}(\mathcal{O}^{f^{*}}, \mathcal{G})$ and  ($mk, vk$)$\leftarrow  \mathbf{KeyGen}()$.
\item  [2.] Compute $\hat{M}\leftarrow \mathbf{FP}(M, mk)$.
\item [3.] Run $\mathcal{A}$ to compute $\tilde{M}\leftarrow \mathcal{A}(\mathcal{O}^f, \hat{M}, vk)$.
\item  [4.] $\mathcal{A}$  wins if
 \begin{equation*}
\begin{split}
&\mathop{Pr}\limits_{x\in {D}}[\mathbf{Classify}(\hat{M}, x)=f^{*}(\mathcal{G}(x))]\approx\\
& \mathop{Pr}\limits_{x\in {D}}[\mathbf{Classify}(\tilde{M}, x)=f^{*}(\mathcal{G}(x))]\ \rm and\; \\
& \mathbf{Verify}(mk, vk, \tilde{M})=0.\\
\end{split}
\end{equation*}
\end{itemize}
Game 1 describes a case where the adversary keeps the first part in the functionality-preserving requirement while violating its second part, which means that the verification samples fail to verify the protected GAN. Therefore, if $\mathcal{A}$ wins Game 1, the fingerprint has been removed. Otherwise, if $\mathcal{A}$ cannot win Game 1, the fingerprint has not been removed, which fulfills the unremovability.

(III) \textbf{Non-rewriteability}.  This property requires that even if the adversary can forge new $mk$ and $vk$ that can be used to pass the verification process, he cannot generate an earlier timestamp than the model owner. Since the generation of the timestamp is equivalent to the time when the model is uploaded to the trusted third party, the adversary cannot forge a timestamp that is earlier than the model owner. Formally, a fingerprinting scheme with non-rewriteability requires that any PPT algorithm $\mathcal{A}$ wins the following \textbf{Game 2} only with negligible probability.
\begin{itemize}
\setlength\itemsep{0.2em}
\item [1.] Generate $M\leftarrow \mathbf{Train}(\mathcal{O}^{f^{*}}, \mathcal{G})$ and ($mk, vk$)$\leftarrow  \mathbf{KeyGen}()$.  
\item  [2.] Compute $\hat{M}\leftarrow \mathbf{FP}(M, mk)$ and generate $ts||s$ binding with $\hat{M}$.
\item [3.] Run the adversary $(\tilde{mk}, \tilde{M}, ts'||s')\leftarrow \mathcal{A}(\mathcal{O}^f, \hat{M}, vk)$.
\item  [4.] $\mathcal{A}$  wins if $\mathbf{Verify}(\tilde{mk}, vk, \tilde{M}$)=1 and $\mathbf{Compare}$($ts||s, c_k, ts'||s', c_k'$)=1. 
\end{itemize}
Game 2 means the adversary creates a new verification protocol for the GAN model with his verification samples and classifier, if needed. $\mathcal{A}$ wins Game 2 if and only if the new verification process passes and its timestamp is earlier. Otherwise, $\mathcal{A}$ fails and the overwritten fingerprint is invalid.

\renewcommand\tablename{Fig.}
\renewcommand \thetable{\arabic{table}}
\setcounter{table}{5}
\begin{table}[h]
\centering
\begin{tabular}{|p{7.8cm}|}
\hline 
 $\mathbf{KeyGen}()$:
\begin{itemize}
\setlength\itemsep{0.2em}
\item [1.] Run  $(V,V_L)=\mathcal{V}\leftarrow\mathbf{F_{gen}}(\mathcal{O}^{f^*}, G)$, where $V=\{v^{(i)}|i\in[n]\}$, $V_L=\{ v_L^{(i)}|i\in[n]\}$.
\item  [2.] Sample  $2n$ random strings $h_v^{(i)}, h_L^{(i)}\leftarrow \{0,1\}^n$, generate $2n$ commitments $\{c_v^{(i)}, c_L^{(i)}\}_{i\in[n]}$, where  $c_v^{(i)}\leftarrow \mathbf{Com}(v^{(i)}, h_v^{(i)})$,  $c_L^{(i)}\leftarrow \mathbf{Com}(v_L^{(i)}, h_L^{(i)})$.
\item  [3.] Set  $mk\leftarrow (\mathcal{V}, \{h_v^{(i)}, h_L^{(i)}\}_{i\in [n]})$, $vk\leftarrow \{c_v^{(i)}, c_L^{(i)}\}_{i\in[n]}$ and return $(mk, vk)$.
\end{itemize}\\
 $\mathbf{FP}$($M, mk$):
\begin{itemize}
\setlength\itemsep{0.2em}
\item [1.] Let $mk=(\mathcal{V}, \{h_v^{(i)}, h_L^{(i)}\}_{i\in [n]})$.
\item  [2.] Compute and output $\hat{M}\leftarrow\mathbf{F_{emb}}(\mathcal{O}^{f}, \mathcal{V}, M)$.
\item [3.] Generate signature keys $p_k$ and $c_k$. Sign the current timestamp $ts$ and obtain $s$.
\item  [4.] Send $G$, $\hat{f}$, $vk$, $ts||s$, and $c_k$ to the trusted third party.
\end{itemize}\\
 $\mathbf{Verify}$($mk, vk, M$):
\begin{itemize}
\setlength\itemsep{0.2em}
\item [1.] Let $mk\leftarrow (\mathcal{V}, \{h_v^{(i)}, h_L^{(i)}\}_{i\in [n]})$, $vk\leftarrow \{c_v^{(i)}, c_L^{(i)}\}_{i\in[n]}$. For  $\mathcal{V}=(V, V_L)$, test if $\forall v^{(i)}\in V$: $ v_L^{(i)}= f({G^s}(v^{(i)}))$. If it is true for all  except $\epsilon|\mathcal{V}|$ elements from $\mathcal{V}$, then output 1, otherwise output 0.
\item  [2.] Check  $\mathbf{Open}$$(c_v^{(i)}, v^{(i)}, h_v^{(i)})=1$ and $\mathbf{Open}$$(c_L^{(i)}, v_L^{(i)}, h_L^{(i)})=1$ \mbox{for all $i\in[n]$. Otherwise, output 0.}
\item [3.] Test  that $\mathbf{Classify}$$(\hat{M}, v^{(i)})=v_L^{(i)}$  for all $i\in[n]$. If it is true for all except $\epsilon|\mathcal{V}|$ elements from $\mathcal{V}$, then output 1, otherwise output 0.
\end{itemize}\\
 $\mathbf{Compare}$($ts||s, c_k, ts'||s', c_k'$):
\begin{itemize}
\setlength\itemsep{0.2em}
\item [1.] Signature check for $ts$, with $s$ and $c_k$. If $ts$ is matched $s$, then output 1, otherwise output 0.
\item  [2.] Signature check for $ts'$, with $s'$ and $c_k'$. If $ts'$ is matched $s'$, then output 1, otherwise output 0.
\item [3.] Order comparison for $ts$ and $ts'$. If $ts$ is earlier than $ts'$, then output 1, otherwise output 0.
\end{itemize}\\
\hline
\end{tabular}
\caption{\textbf{End-to-end fingerprinting process.}}
\label{Concrete Construction of fingerprinting process}
\vspace{-15pt}
\end{table}

\renewcommand{\tablename}{Table}
\setcounter{table}{1}

\vspace{-5pt}
\section{Concrete Methodologies of Generating and Embedding Strong Fingerprints}
\label{Concrete Fingerprinting Construction}
\vspace{-5pt}
\subsection{Assumptions for Strong Fingerprints}
\label{Strong Fingerprints}
\vspace{-5pt}

With the two algorithms $\mathbf{F_{gen}}$ and $\mathbf{F_{emb}}$, we expect that the model owner can produce strong fingerprints $\mathcal{V}$ obeying three properties. $\mathbf{F_{emb}}$ that takes such samples as input is called a strong fingerprinting algorithm. These are necessary for us to build effective fingerprinting solutions that meet the requirements in Section \ref{security requirements}.


(1) \textit{Stealthiness}:  Each verification sample during inference should be indistinguishable from the normal ones, making it difficult for the adversary to respond adaptively and ensuring the concealment of verification. This means that for each verification sample $v^{(i)}\in V$ generated from a randomly selected clean sample $x^{(i)}$, the following expression:
\begin{align*}
    \mathcal{H} =\Vert v^{(i)}, x^{(i)} \Vert + \Vert G(v^{(i)}), G(x^{(i)}) \Vert \\
    + \sum_j \Vert G_{v^{(i)},j}, G_{x^{(i)},j} \Vert
\end{align*}
is minimized, where $G_{v^{(i)},j}$ is the $j$-th intermediate feature in $G$ with input $v^{(i)}$, $G(v^{(i)})$ is the output of $G$ with input $v^{(i)}$, and $\Vert \cdot, \cdot \Vert$ is a distance function\footnote{Note that the stealthiness of fingerprints is difficult to describe with cryptographic primitives because it is very subjective. We mainly demonstrate this property based on empirical experiments.}. The first term describes the differences between the verification sample and the corresponding clean image in the pixel space. The second term is to measure the distance between the output of the verification sample and that of the corresponding clean sample in the pixel space. The third term is to compare the intermediate features in $G$. Verification samples achieve stealthiness in three levels.

(2) \textit{Persistency}: $\mathbf{F_{emb}}$ is able to embed the fingerprint persistently such that the adversary cannot remove the fingerprint from the model. This property is discussed under two assumptions. First, the adversary has limited computing resources and data resources, which do not support him to retrain a clean model with competitive performance from scratch. Otherwise, he will lose the motivation of stealing others' models. Second, the adversary is not willing to erase the fingerprint at the cost of huge accuracy drop for the plagiarized model. Hence, we define the persistency as follows: let $\mathcal{O}^{f^*}$ be a   ground-truth oracle, $\mathcal{V}$  be a fingerprint set, and  $\hat{M}\leftarrow\mathbf{F_{emb}}(\mathcal{O}^{f^*}, \mathcal{V}, M)$ be a $\epsilon$-accurate model.   Assume  an algorithm $\mathcal{A}$ on input $\mathcal{O}^{f^*}$,  $\hat{M}$  outputs a model $\tilde{M}$  in polynomial time $t$ which  is at least $(1-\epsilon)$ accurate on $\mathcal{V}$. Then, for any arbitrary model $N$,  $\tilde{N}\leftarrow$$\mathbf{F_{emb}}$$(\mathcal{O}^{f^*}, N)$ generated in same time $t$, is also $\epsilon$-accurate (related to Game 1). 

Below we present three novel concrete designs based on our fingerprinting scheme. For each design, we describe the two crucial algorithms $\mathbf{F_{gen}}$ and $\mathbf{F_{emb}}$ for generating strong fingerprints and embedding them into the model, respectively.

\vspace{-5pt}
\subsection{\Namea}
\label{sec:design1}
\vspace{-5pt}

\renewcommand{\algorithmicrequire}{$\mathbf{F_{gen}}$ ($\mathcal{O}^{f^{*}}, G$)}
\begin{algorithm}
\caption{Fingerprint Generation}
\label{algorithm 1}
\begin{algorithmic}[1]
\REQUIRE
\STATE Train a normal classifier $f$ with $\mathcal{O}^{f^{*}}$ and  target GAN model $G$.
\STATE Uniformly select random samples $\{x,y\} \in \bar{D}$  $n$ times to build $X =\{x^{(1)}, \cdots, x^{(n)}\} $ and $Y = \{y^{(1)}, \cdots, y^{(n)}\}$.
\FOR{each $\{x^{(i)}, y^{(i)}\} \in \{X, Y\}$}
\STATE Generate $v^{(i)}$ from $\{x^{(i)}, y^{(i)}\}$ by minimizing the objective function $F_{obj}(\mathcal{O}^{f},G, \{x^{(i)}, y^{(i)}\}, v^{(i)})$ in Equation \ref{equ:loss1}. 
\STATE Generate $\{v_L^{(i)}|i\in[n]\}$ with label $v_L^{(i)} = f(G(v^{(i)})) \neq f^{*}(G(v^{(i)}))$.
\ENDFOR
\STATE \textbf{Return} a fingerprint $\mathcal{V}=(V,V_L)$, where $V=\{v^{(i)}|i\in[n]\}$ and $V_L=\{ v_L^{(i)}|i\in[n]\}$.
\end{algorithmic}
\end{algorithm}

Our first method, \Namea (Composite Fingerprint based on Adversarial Examples), is inspired by the generative adversarial examples \cite{xiao2018generating}. Different from the traditional fingerprinting methods \cite{cao2019ipguard,lukas2019deep} that directly craft adversarial examples against the target model, we propose to make the target GAN model generate adversarial examples to the classifier. The output sample of the target model looks normal, while the output {label of the classifier is unique as the ownership evidence.} 

Algorithm~\ref{algorithm 1} shows the detailed process of generating the fingerprint set. Given $\mathcal{O}^{f^*}$,we first train a classifier or download a pre-trained classifier $f$ for classifying the attributes or categories of the data samples. Then we uniformly select some random samples $\{x^{(i)}| {i\in[n]}\}$ from the sample space $D$. Since the size of the space is $\Theta (2^n)$, a PPT adversary only has a negligible probability to infer these samples. We craft the verification samples $v^{(i)}$ from these clean samples using an optimization method, by optimizing the input to minimize an object function. To ensure the indistinguishability between the verification sample and its corresponding clean sample, for each $v^{(i)}$,  we need to minimize $\mathcal{H} =\Vert v^{(i)}, x^{(i)} \Vert + \Vert G(v^{(i)}), G(x^{(i)}) \Vert + \sum_j \Vert G_{v^{(i)},j}, G_{x^{(i)},j} \Vert$.
Also, a qualified verification sample $v^{(i)}$ should enable the classifier to maximize the distance between the ground-truth label $y^{(i)}$ corresponding to $G(x^{(i)})$ and predicted label $f(G(v^{(i)}))$. To achieve these,  we construct a loss function  $F_{obj}(\mathcal{O}^{f^{}},G, \{x^{(i)}, y^{(i)}\}, v^{(i)})$ as:
\begin{equation}
\label{equ:loss1}
\begin{aligned}
 F_{obj}(&\mathcal{O}^{f^{}},  G, \{x^{(i)}, y^{(i)}\}, v^{(i)}) = \\
 & \sum_{c} y^{(i)}_{c}\log(f(G(v^{(i)}))_{c}) +\sum (v^{(i)}- x^{(i)})^2 +  \\
 & \sum (G(v^{(i)}) - G(x^{(i)}))^2 +\\
 & \sum_j \sum (G_{v^{(i)},j} - G_{x^{(i)},j})^2,
\end{aligned}
\end{equation}
where $G_{v^{(i)},j}$ and $G_{x^{(i)},j}$ are the $j$-th features in $G$ when processing $v^{(i)}$ and $x^{(i)}$, respectively. The subscript $c$ stands for the index of the ground-truth label, which is converted to a one-hot vector, and the prediction of the classifier. The first term in the object function is based on the cross-entropy loss by multiplying -1 to maximize the difference between the prediction and the ground-truth label. We iteratively search for the optimal $v^{(i)}$ by minimizing the above objective function. As a result, we  obtain the final verification sample $v^{(i)}$ with the label $f(G(v^{(i)}))=v_L^{(i)}\neq f^{*}(G(v^{(i)}))$.

It is worth noting that in \Namea, we do not need to modify the classifier $f$ after we perform the $\mathbf{F_{gen}}$ function. We directly use the generated samples to query the composite model for ownership verification. Hence, the $\mathbf{F_{emb}}$ function is empty with $\hat{f}=f$ in this method.

\vspace{-5pt}
\subsection{\Nameb}
\label{sec:design2}
\vspace{-5pt}

Our second method, \Nameb (Composite Fingerprint based on invisible Backdoor (version 1)), utilizes the invisible backdoor attack technique \cite{li2019invisible}. The key idea is to make the target model produce output samples containing invisible triggers, which will activate the backdoor embedded in the classifier to predict unique labels. 
\Nameb requires two steps. Fingerprint generation calls the same function $\mathbf{F_{gen}}$ as in \Namea to produce verification samples and labels. Then we perform the fingerprint embedding $\mathbf{F_{emb}}(\mathcal{O}^{f}, \mathcal{V}, M)$, which further fine-tunes the classifier $f$ into $\hat{f}$, to better recognize the relationships between the verification samples and labels.

Algorithm \ref{algorithm 2} shows the detailed process of fine-tuning the classifier. We prepare two sets: the verification set $\mathcal{V}_s = (G(V), V_L)$, where $V$ and $V_L$ are generated from $\mathbf{F_{gen}}$; the normal set $\mathcal{N}_s = (G(X), Y)$, where $X$ contains samples generating $V$ in $\mathbf{F_{gen}}$ and $Y$ contains labels corresonding to samples in $G(X)$. Since the fingerprint must be persistent against image transformations, we further perform data augmentation over these two sets with common transformation functions. Using these two augmented sets $\mathcal{V}^a_s$ and $\mathcal{N}^a_s$, we fine-tune the classifier as $\hat{f}$, and finally obtain the composite model $\hat{M}(\cdot)=\hat{f}({G}(\cdot))$.

\renewcommand{\algorithmicrequire}{$\mathbf{F_{emb}}(\mathcal{O}^{f}, \mathcal{V}, M)$}
\begin{algorithm}
\caption{Fingerprint Embedding}
\label{algorithm 2}
\begin{algorithmic}[1]
\REQUIRE
\STATE $(V, V_L) = \mathcal{V}$.
\STATE $\mathcal{V}_s = (G(V), V_L)$.
\STATE $\mathcal{N}_s = (G(X), Y)$, $X$ and $Y$ are from $\mathbf{F_{gen}}$.
\STATE Augment these two sets to obtain $\mathcal{V}^a_s$ and $\mathcal{N}^a_s$.
\STATE Fine-tune $f$ into $\hat{f}$ with $\mathcal{V}^a_s$ and $\mathcal{N}^a_s$ together by minimizing the loss function $\mathcal{L}_{ft}$.
\STATE \textbf{Return} fingerprinted model $\hat{M}(\cdot)=\hat{f}({G}(\cdot))$.
\end{algorithmic}
\end{algorithm}

We use the cross-entropy loss function to fine-tune the classifier with the two sets:
\begin{align*}
    \mathcal{L}_{ft} = \mathcal{L}_{G1}(\mathcal{O}^{f}, \mathcal{V}^a_s, \mathcal{N}^a_s) =&-\sum_{ (x,y)\in\mathcal{V}^a_s}\sum_c y_{c}\log(f(x)_{c}) \\
    &-\sum_{ (x,y)\in\mathcal{N}^a_s}\sum_c y_{c}\log(f(x)_{c}),
\end{align*}
where $c$ is the label index of $f$. In the fine-tuning loss, we aim to minimize the cross-entropy loss on two sets, $\mathcal{N}^a_s$ and $\mathcal{V}^a_s$, and make the fine-tuned classifier $\hat{f}$ sensitive to the small difference between $G(x)$ and $G(v)$ to give them different predictions.

\vspace{-5pt}
\subsection{\Name}
\label{sec:design3}
\vspace{-5pt}

Our third method, \Name (Composite Fingerprint based on invisible Backdoor (version 2)), is an advanced version of \Nameb. We follow the same algorithms to generate fingerprints and embed them into the model. A novel loss function is introduced to fine-tune the classifier for better robustness and effectiveness. 

First, we adopt the idea of the Triplet Loss~\cite{schroff_facenet_2015} to enhance the persistency of our fingerprints. The Triplet Loss is able to distinguish different objects under similar conditions (e.g., pose, illumination). It achieves this by minimizing the inner representation (i.e., feature embedding) difference of the same object with different external conditions, while maximizing the difference of different objects with the same condition. Similarly, we can minimize the distance of different verification samples in the feature space, and maximize the distance of a verification and normal samples. This can increase the probability that the fine-tuned classifier will give unique labels for verification samples. The loss function is as below: 
\begin{align*}\nonumber
     \mathcal{L}_{G2}(\mathcal{M},\mathcal{V}^a_s, &\mathcal{N}^a_s, m) = \\
    & \sum_{v_a\in \mathcal{V}^a_s} \max\{ \max_{v_p\in \mathcal{V}^a_s}(\sum (\mathcal{M}(v_a) - \mathcal{M}(v_p))^2) \\
    &- \min_{x\in \mathcal{N}^a_s}( \sum (\mathcal{M}(v_a) - \mathcal{M}(x))^2 ) +m ,0\}, 
\end{align*}
where $m$ is a constant, and $\mathcal{M}(\cdot)$ represents the feature extraction part in $f$ before the final classification layer. $v_a$ and $v_p$ are from $\mathcal{V}^a_s$ and $x$ is from $\mathcal{N}^a_s$. $v_a$ is an anchor sample, $v_p$ is the positive sample, and $x$ is the negative sample. The goal of the Triplet Loss is to minimize the distance of features between the anchor sample and the positive sample and maximize the distance of features between the anchor sample and the negative sample. By minimizing this loss, we can make $f$ assign similar features to $G(v)$ for all verification samples, which will be very different from the features of $G(x)$. Therefore, $\hat{f}$ will be more robust to recognize $G(v)$, making the verification process more reliable.

Second, we apply the fine-grained categorization approaches \cite{deng2013fine,gavves2013fine} to fine-tune the classifier. Fine-grained categorization aims to classify an object into an exact sub-category, e.g., the brand of a car, the species of a bird. Various techniques have been introduced to achieve this challenging goal~\cite{luo_cross-x_2019,yang_learning_2018,  zhang_multi-branch_2021}. We can treat the fingerprint embedding process as a fine-grained categorization task, 
where samples from $\mathcal{V}^a_s$ are in one category (fingerprint verification), while samples from $\mathcal{N}^a_{s}$ are in another category (normal inference). Specifically, we change the classifier to a multitask one by adding an additional classification head to the original model structure: the original classification layer is used to predict the category labels for GAN's output, while a new one is added to predict the verification category (label ``1'' for fingerprint verification; label ``0'' for normal inference). Then we adopt the Entropy-Confusion Loss~\cite{dubey_maximum-entropy_2018} to train the multitask model:
\begin{align*}
   \mathcal{L}_{G3}(& \mathcal{B},  \mathcal{V}^a_s, \mathcal{N}^a_s, \epsilon)  = \\
    & \sum_{v\in \mathcal{V}^a_s}(\mathcal{B}(v)_0 \log{\frac{\mathcal{B}(v)_0}{\mathcal{B}(v)_1 + \epsilon}} + (\mathcal{B}(v)_1 + 1) \log{\mathcal{B}(v)_1}) + \\
    & \sum_{x\in \mathcal{N}^a_s}((\mathcal{B}(x)_0 + 1) \log{\mathcal{B}(x)_0} + \mathcal{B}(x)_1 \log{\frac{\mathcal{B}(x)_1}{\mathcal{B}(v)_0 + \epsilon}}),
\end{align*}
where $\epsilon = 1e^{-5}$ is a constant to avoid a denominator of zero, $\mathcal{B}(\cdot)$ is the output from the added binary classification layer, and $\mathcal{B}(\cdot)_i$ is the $i$-th element in the output. We force the new classification layer to give $G(v)$ the prediction of the verification sample and give $G(x)$ the prediction of the normal data. In this way, we further improve $f$'s robustness in recognizing $G(v)$ and $G(x)$. Hence, the ultimate loss $\mathcal{L}_{ft}$ used to fine-tune $f$ is
\begin{align*}
            \mathcal{L}_{ft}= \mathcal{L}_{G1} + \mathcal{L}_{G2} + \mathcal{L}_{G3}.
\end{align*}
After we finish the classifier fine-tuning, we remove the binary classification layer from $\hat{f}$, and integrate it with the target GAN model to form the composite model $\hat{M}$. 

\vspace{-5pt}
\subsection{Generalization to Different I2I Tasks}
\vspace{-5pt}

Our proposed fingerprinting schemes are suitable for different I2I GAN models and tasks. Below we describe a unified process for general I2I GANs. First, we need to build a classification model for verification sample generation. The choice of the classifier is flexible, and we only ask it to assign a label to the given image, which can be unrelated to the image content. For example, for all I2I GANs, the classifier can be trained on ImageNet. In our experiments, to better show the choice of the classifier is flexible, we choose different classifiers for each task.

Then, we need to generate some verification samples for the protected GAN with the selected classifier. Considering that the training data for generative models usually do not contain a label, we directly use the predicted label from the classifier $f$ as the ground-truth label $y$ for the clean data $x$. That is why our fingerprinting scheme does not require a specific classifier. Therefore, all chosen $f$ can be seen as $f^*$, outputting ground-truth labels. With the label $y$ and the classifier $f$, we can generate each verification sample $v$ based on $x$ for the protected model $G$. Finally, we can fine-tune $f$ to obtain $\hat{f}$. Therefore, our scheme is a unified protection for different tasks.

\vspace{-5pt}
\section{Security Analysis}
\label{sec:sec-analysis}
\vspace{-5pt}

Assuming $\mathbf{F_{emb}}$ is a strong fingerprinting algorithm that can generate fingerprints with the three properties in Section \ref{Strong Fingerprints}, we prove our fingerprinting scheme can satisfy the three requirements in Section \ref{security requirements} in the following theory. Therefore, we build a connection between security requirements and our proposed fingerprinting scheme, proving it is an effective solution.

\begin{theorem}
Let $\bar{D}$ be of super-polynomial size in $n$. Given the commitment scheme and the strong fingerprinting algorithm, the algorithms ($\mathbf{KeyGen}$, $\mathbf{FP}$, $\mathbf{Verify}$) in Fig.~\ref{Concrete Construction of fingerprinting process} form a privately verifiable fingerprinting scheme, which satisfies the  requirements of functionality-preserving, unremovability, and non-rewriteability.
\end{theorem}
\noindent\textit{Proof (Sketch)}: 
Our proof relies on the security of the commitment scheme and assumptions of strong fingerprints. In detail, the \textit{hiding} property of the commitment scheme enables the public verification key to hide useful information about the fingerprint from the adversary, while the \textit{binding} property ensures that one cannot claim the ownership of a model from others. Also, as defined in Section~\ref{Strong Fingerprints}, a strong fingerprint  embedded into the model should (i) be distinctive, i.e., behaves like $f^{*}(\mathcal{G}(\cdot))$ on $\bar{D}\backslash V$ and reliably predicts unique labels $V_L$ on $V$ (i.e., functionality-preserving), and (ii) be persistent such that the adversary cannot revert the fingerprinted model back to the original one in time $t$ (i.e.,  unremovability). In addition, since the generation of the timestamp is equivalent to the time when the model is uploaded to the trusted third party, it is impossible for the adversary to forge a timestamp that is earlier than the model owner (non-rewriteability). Therefore, we can prove that our designs satisfy the above requirements. \textbf{For the technical \mbox{details of the proof, please see Appendix~\ref{sec:proofs}.}}

\textit{Remark}: The algorithm $\mathbf{Verify}$ only allows verification by honest parties in a private way, since  $mk$ will be known once $\mathbf{Verify}$ is run, which allows the adversary to retrain the model on the verification sample set. It is not a problem for the applications such as IP protection, because there are trusted third parties in the form of judges. 

\vspace{-5pt}
\section{Experiments}
\label{sec:exp}
\vspace{-5pt}

We conduct comprehensive experiments to validate that our concrete designs can meet the strong fingerprint requirements in Section~\ref{Strong Fingerprints}. We report the main results in this section, and a plethora of experimental results can be found in Appendix~\ref{appendix:pruning}.

\renewcommand\tablename{Table}
\noindent\textbf{Model and dataset.} Our scheme can be applied to general I2I GAN models and tasks, since the design does not rely on any assumptions about datasets, model architectures or parameters. Without loss of generality, we evaluate GANs for three I2I tasks, i.e., attribute editing, domain translation, and super resolution, with various GAN models. Specifically, for attribute editing, we train three GANs (AttGAN~\cite{he_attgan_2019}, StarGAN~\cite{choi_stargan_2018}, STGAN~\cite{liu_stgan_2019}) to edit five attributes: (A1) black hair, (A2) blond hair, (A3) brown hair, (A4) male, and (A5) young, on a public dataset CelebA~\cite{liu_deep_2015}. For domain translation, we train three CycleGANs~\cite{cyclegan}, with different batch sizes and random seeds, named C1, C2, and C3 to achieve a horse-to-zebra task~\cite{cyclegan}. For super resolution, we train three GANs (SRResNet~\cite{srresnet}, ESRGAN~\cite{esrgan}, EDSR~\cite{edsr}) on DIV2K~\cite{div2k} to achieve a 2$\times$ up-scaling super-resolution.  Our main experiments are conducted on the attribute editing task. For the other two tasks, we use them to evaluate the generalizability of our schemes.

\noindent\textbf{Scheme implementation.} 
In our experiments, the classifier $f$ is implemented by ResNet34~\cite{he_deep_2016}. For attribute editing, we train $f$ as a multi-label classifier on the CelebA dataset to predict the facial attributes. Each sample in CelebA has $40$ annotated attributes. Then the output of $f$ is a 40-bit vector, with each bit representing whether the image has the corresponding attribute. For domain translation, we train $f$ as a two-class classifier to recognize horses and zebras. For super resolution, we train $f$ on ImageNet as a classifier to recognize 1,000 categories. Note that the construction of $f$ is general, so other mainstream classification models can be applied to our tasks as well.

For $\mathbf{F_{gen}}$ in Algorithm~\ref{algorithm 1}, we select 100 random images as clean data set $X$ to generate the verification sample set $V$. Specifically, for attribute editing, images are selected from CelebA. For domain translation, images are all horses. For super resolution, images are selected from ImageNet. For each sample, we set its unique verification label $V_L(x)$ after generating the verification sample by minimizing $F_{obj}$ in Equation~\ref{equ:loss1}. Then, $V_L(x)$ is determined by $f$ based on the prediction of the verification sample. We set the optimization constraint $\Vert G(v^{(i)}), G(x^{(i)}) \Vert \leq \delta=9e^{-4}$, which is proven to be sufficient to ensure the indistinguishability between the verification sample and its corresponding clean sample. The generated verification sample set can be used for all three proposed methods. A slight difference between attribute editing and other tasks is that after $V$ is crafted, we keep all the flipped 40 attributes as the verification label for \Nameb and \Name. For \Namea, we only flip 5 attributes, while the rest attributes are the same as the ground truth. These 5 attributes are selected as the easiest to be misclassified by analyzing the decision boundary of the classifier. Table~\ref{tab:attributes} shows these attributes for each GAN model. For other tasks, the verification label is a single number. The difference is mainly because attribute editing GANs could influence multiple attributes, which could cause mis-verification when only adopting one attribute as the verification label.

\setcounter{table}{0}
\renewcommand{\arraystretch}{1.2}
\begin{table}[h]
\centering
\caption{\textbf{Top-$5$ attributes for three GANs in verifying the model with} \Namea.}
\begin{adjustbox}{max width=\linewidth}
\begin{tabular}{c|c}
\Xhline{1pt}
\textbf{GAN}         & \textbf{Selected fingerprinting attributes} \\
\Xhline{1pt}
AttGAN & Smiling, BagsUnderEyes, Attractive, MouthSlightlyOpen, HighCheekbones  \\ \hline
StarGAN & Smiling, Male, Young, WearingNecklace, Attractive  \\ \hline
STGAN & BigNose, Young, Smiling, BagsUnderEyes, HighCheekbones\\
\Xhline{1pt}
\end{tabular}
\end{adjustbox}
\vspace{-5pt}
\label{tab:attributes}
\end{table}

For \Namea, we do not need to make any changes to the classifier $f$. For \Nameb and \Name, we need to embed the fingerprint into the composite model following $\mathbf{F_{emb}}$ in Algorithm~\ref{algorithm 2}. We fix $\mathcal{G}$, while fine-tuning the classifier $f$ using the prepared verification sample set.  This will give us the final fingerprint-embedded composite model $\hat{M}=\hat{f}({G}(\cdot))$. 
For $\mathbf{F_{emb}}$ in Algorithm~\ref{algorithm 2},
to enhance the robustness of the fingerprinted classifier, we adopt four types of mainstream image transformations (adding noise, blurring, compression and cropping) to augment the verification sample set $\mathcal{V}_s$ and normal sample set $\mathcal{N}_s$. We fine-tune $f$ with only 100 verification samples and 100 normal samples, so it is very efficient for the model owner to annotate these samples.

For verification, we query the suspicious GAN model with $100$ verification samples. Similar to prior works \cite{le2019adversarial,li2019prove,cao2019ipguard,wang2021fingerprinting,lukas2019deep}, we empirically set the threshold $\tau$ for ownership judgement, which is 0.8. 

\setcounter{table}{1}
\begin{table}[ht]
\centering
\caption{\textbf{MSC (\%) and MSV (\%) for verifying different GAN models}. $\Downarrow$ means a lower score is better. $\Uparrow$ means a higher score is better. Same for the following tables.}
\begin{adjustbox}{max width=\linewidth}
\begin{tabular}{c|c|cc|ccc}
\Xhline{1pt}
\multirow{3}{*}{\textbf{\begin{tabular}[c]{@{}c@{}}GAN\\ Structure\end{tabular}}} & \multirow{3}{*}{\textbf{Method}} & \multicolumn{2}{c|}{\textbf{Target GAN}} & \multicolumn{3}{c}{\textbf{Non-target GAN}} \\ \cline{3-7} 
 &  & \multirow{2}{*}{MSC $\Uparrow$} & \multirow{2}{*}{MSV $\Uparrow$} & \multicolumn{1}{c|}{StarGAN} & \multicolumn{1}{c|}{AttGAN} & STGAN \\ \cline{5-7} 
 &  &  &  & \multicolumn{1}{c|}{MSV $\Downarrow$} & \multicolumn{1}{c|}{MSV $\Downarrow$} & MSV $\Downarrow$ \\ \hline
\multirow{5}{*}{StarGAN} & \texttt{AE-I} & 100.00 & 100.00 & \multicolumn{1}{c|}{0.00} & \multicolumn{1}{c|}{0.00} & 0.00 \\
 & \texttt{AE-D} & 100.00 & 100.00 & \multicolumn{1}{c|}{100.00} & \multicolumn{1}{c|}{20.00} & 12.00 \\
 & \texttt{CFP-AE} & 100.00 & 100.00 & \multicolumn{1}{c|}{50.20} & \multicolumn{1}{c|}{33.80} & 30.40 \\
 & \texttt{CFP-iBDv1} & 95.52 & 94.12 & \multicolumn{1}{c|}{62.10} & \multicolumn{1}{c|}{15.10} & 27.00 \\
 & \texttt{CFP-iBDv2} & 92.87 & 90.05 & \multicolumn{1}{c|}{39.62} & \multicolumn{1}{c|}{12.53} & 16.92 \\ \hline
\multirow{5}{*}{AttGAN} & \texttt{AE-I} & 100.00 & 100.00 & \multicolumn{1}{c|}{0.00} & \multicolumn{1}{c|}{0.00} & 0.00 \\
 & \texttt{AE-D} & 100.00 & 14.00 & \multicolumn{1}{c|}{35.00} & \multicolumn{1}{c|}{1.00} & 6.00 \\
 & \texttt{CFP-AE} & 100.00 & 100.00 & \multicolumn{1}{c|}{29.00} & \multicolumn{1}{c|}{42.20} & 39.80 \\
 & \texttt{CFP-iBDv1} & 93.40 & 92.45 & \multicolumn{1}{c|}{49.10} & \multicolumn{1}{c|}{34.20} & 57.02 \\
 & \texttt{CFP-iBDv2} & 91.03 & 90.70 & \multicolumn{1}{c|}{27.15} & \multicolumn{1}{c|}{17.85} & 30.10 \\ \hline
\multirow{5}{*}{STGAN} & \texttt{AE-I} & 98.00 & 100.00 & \multicolumn{1}{c|}{0.00} & \multicolumn{1}{c|}{0.00} & 13.00 \\
 & \texttt{AE-D} & 100.00 & 34.00 & \multicolumn{1}{c|}{66.00} & \multicolumn{1}{c|}{22.00} & 1.00 \\
 & \texttt{CFP-AE} & 100.00 & 100.00 & \multicolumn{1}{c|}{26.20} & \multicolumn{1}{c|}{25.80} & 67.80 \\
 & \texttt{CFP-iBDv1} & 93.53 & 91.57 & \multicolumn{1}{c|}{50.52} & \multicolumn{1}{c|}{42.08} & 83.20 \\
 & \texttt{CFP-iBDv2} & 92.20 & 90.18 & \multicolumn{1}{c|}{30.05} & \multicolumn{1}{c|}{28.75} & 69.05\\
\Xhline{1pt}
\end{tabular}
\end{adjustbox}
\vspace{-5pt}
\label{tab:otherganandclassifier}
\end{table}

\noindent\textbf{Baselines.} 
Since there are no existing works for fingerprinting I2I GAN models, we migrate the fingerprinting strategy from classification models to GANs as our baselines. Past works proposed two types of common techniques to generate adversarial attacks for GAN models, which are adopted for fingerprint generation in our baselines. Specifically, (1) \texttt{AE-D} leverages the \textit{distortion attack} \cite{aed3,aed2,aed,aes2,aes}, whose outputs are distorted away from the correct one. This is achieved by maximizing the distance between the adversarial output and ground-truth output. During verification, we determine the legitimacy of the suspicious model by measuring the noise ratio of the responses and ground-truth outputs. A model is considered as illegal if the peak signal-to-noise ratio (PSNR)~\cite{psnrssim} is smaller than a threshold ($20$). (2) \texttt{AE-I} leverages the \textit{identity attack}~\cite{aes2,aes}, whose outputs are identical with the inputs. This is achieved by minimizing the distance between the sample outputs and inputs. During verification, we measure the similarity between the verification samples and the corresponding responses. We flag the model as pirated if their Euclidean distance is smaller than a threshold ($9e^{-4}$). Both types of adversarial examples are generated by C\&W~\cite{carlini2017towards}, which is also used in~\cite{cao2019ipguard} for fingerprinting classification models.

\noindent\textbf{Metrics.}
We introduce two metrics: (1) Match Score for Verification samples (MSV) denotes the match ratio of verification labels for verification samples; (2) Match Score for Clean samples (MSC) denotes the match ratio of ground-truth labels for clean samples. For a good fingerprinting method, the target model should have high MSV and MSC, while the MSV on unrelated models should be low. 

\vspace{-5pt}
\subsection{Time Cost Comparison}\label{sec:time}
\vspace{-5pt}

Training a high-quality GAN will cost a lot of time. For example, training a StarGAN~\cite{choi_stargan_2018}, used in our paper, on one V100 will cost about one week to achieve good performance. Training a StyleGAN~\cite{stylegan}, which is a popular generative model, on 8 GPUs will cost one week to generate high-resolution images\footnote{https://github.com/NVlabs/stylegan}. Compared with the training cost, generating one image as a fingerprint to verify the GAN only needs several minutes, depending on the GAN itself. Therefore, our protection scheme is efficient and environmentally friendly.

\vspace{-5pt}
\subsection{Distinctness  Analysis}
\label{sec:unique-analysis}
\vspace{-5pt}

We show that the generated verification samples can identify the target GAN models with a higher probability. We generate verification samples and fingerprinted classifier from one target GAN model, and use them to verify the model itself, as well as other unrelated GAN models, including a model trained with the same configurations (network structure, algorithm, hyperparameters and dataset). Table~\ref{tab:otherganandclassifier} presents the Match Scores for different models. We observe that all methods perform well on the target model. For other unrelated models, \texttt{AE-I} performs the best in reducing the false positives. This indicates the adversarial identity attack has much lower transferability to other models. We will show that \texttt{AE-I} is impractical in terms of persistency (Section~\ref{sec:robust-analysis}). \texttt{AE-D} has high transferability for StarGAN, hence it fails to distinguish target and non-target GAN models trained from the same StarGAN. Our methods are generally fair to distinguish target and non-target models with a threshold $\tau=0.8$. \texttt{CFP-iBDv2} is better than \texttt{CFP-AE} and \texttt{CFP-iBDv1}, due to the utilization of more sophisticated loss functions when fine-tuning the classifier. 

\vspace{-5pt}
\subsection{Persistency Analysis}
\label{sec:robust-analysis}
\vspace{-5pt}

We adopt the mainstream operations in prior watermarking or fingerprinting works \cite{adi2018turning,le2019adversarial,li2019prove,zhang2018protecting,cao2019ipguard,lukas2019deep,wang2021characteristic,wang2021fingerprinting} to evaluate the persistency of different fingerprinting methods. 

\begin{table*}[t]
\scriptsize
\centering
\caption{\textbf{MSC (\%) and MSV (\%) after two model transformations}.}
\begin{adjustbox}{max width=0.95\linewidth}
\begin{tabular}{c|c|cc|cc|cc|cc|cc|cc}
 \Xhline{1pt}
\multirow{3}{*}{\textbf{\begin{tabular}[c]{@{}c@{}}GAN\\ Structure\end{tabular}}} & \multirow{3}{*}{\textbf{Method}} & \multicolumn{2}{c|}{\multirow{2}{*}{\textbf{Target GAN}}} & \multicolumn{6}{c|}{\textbf{Fine-tuning (epochs)}} & \multicolumn{4}{c}{\textbf{Pruning (compression ratio)}} \\ \cline{5-14} 
 &  & \multicolumn{2}{c|}{} & \multicolumn{2}{c|}{10} & \multicolumn{2}{c|}{20} & \multicolumn{2}{c|}{30} & \multicolumn{2}{c|}{0.2} & \multicolumn{2}{c}{0.4} \\ \cline{3-14} 
 &  & MSC $\Uparrow$ & MSV $\Uparrow$ & MSC $\Uparrow$ & MSV $\Uparrow$ & MSC $\Uparrow$ & MSV $\Uparrow$ & MSC $\Uparrow$ & MSV $\Uparrow$ & MSC $\Uparrow$ & MSV $\Uparrow$ & MSC $\Uparrow$ & MSV $\Uparrow$ \\ \hline
\multirow{5}{*}{StarGAN} & \texttt{AE-I} & 100.00 & 100.00 & 100.00 & 0.00 & 100.00 & 6.00 & 100.00 & 2.00 & 100.00 & 43.00 & 100.00 & 0.00 \\
 & \texttt{AE-D} & 100.00 & 100.00 & 100.00 & 100.00 & 99.00 & 100.00 & 100.00 & 100.00 & 100.00 & 100.00 & 100.00 & 100.00 \\
 & \texttt{CFP-AE} & 100.00 & 100.00 & 96.60 & 94.20 & 96.40 & 91.80 & 97.00 & 96.00 & 98.80 & 98.20 & 94.60 & 97.00 \\
 & \texttt{CFP-iBDv1} & 95.52 & 94.12 & 94.98 & 93.07 & 92.20 & 89.25 & 92.88 & 92.15 & 95.58 & 94.12 & 95.35 & 93.60 \\
 & \texttt{CFP-iBDv2} & 92.87 & 90.05 & 92.53 & 85.32 & 92.28 & 84.57 & 92.68 & 84.80 & 92.73 & 90.02 & 92.78 & 89.57 \\ \hline
\multirow{5}{*}{AttGAN} & \texttt{AE-I} & 100.00 & 100.00 & 100.00 & 91.00 & 100.00 & 84.00 & 100.00 & 75.00 & 100.00 & 22.00 & 100.00 & 0.00 \\
 & \texttt{AE-D} & 100.00 & 14.00 & 100.00 & 14.00 & 100.00 & 14.00 & 100.00 & 14.00 & 100.00 & 14.00 & 100.00 & 16.00 \\
 & \texttt{CFP-AE} & 100.00 & 100.00 & 98.60 & 94.60 & 99.80 & 95.40 & 99.00 & 94.80 & 97.80 & 91.20 & 86.40 & 87.40 \\
 & \texttt{CFP-iBDv1} & 93.40 & 92.45 & 93.33 & 92.37 & 93.45 & 92.40 & 93.45 & 92.37 & 93.53 & 92.40 & 93.08 & 89.00 \\
 & \texttt{CFP-iBDv2} & 91.03 & 90.70 & 91.93 & 90.62 & 92.00 & 90.70 & 92.05 & 90.67 & 91.98 & 90.75 & 91.95 & 84.95 \\ \hline
\multirow{5}{*}{STGAN} & \texttt{AE-I} & 98.00 & 100.00 & 100.00 & 85.00 & 99.00 & 75.00 & 92.00 & 73.00 & 100.00 & 58.00 & 100.00 & 0.00 \\
 & \texttt{AE-D} & 100.00 & 34.00 & 100.00 & 36.00 & 100.00 & 36.00 & 100.00 & 32.00 & 100.00 & 34.00 & 100.00 & 57.00 \\
 & \texttt{CFP-AE} & 100.00 & 100.00 & 99.40 & 95.40 & 99.80 & 95.20 & 99.40 & 94.60 & 98.60 & 95.00 & 93.40 & 86.60 \\
 & \texttt{CFP-iBDv1} & 93.53 & 91.57 & 93.58 & 91.62 & 93.53 & 91.72 & 93.28 & 91.80 & 93.38 & 91.45 & 84.20 & 91.40 \\
 & \texttt{CFP-iBDv2} & 92.20 & 90.18 & 92.08 & 90.30 & 92.20 & 90.25 & 91.95 & 90.40 & 91.55 & 90.22 & 88.98 & 83.35 \\
 \Xhline{1pt}
\end{tabular}
\end{adjustbox}
\label{tab:robust}
\vspace{-5pt}
\end{table*}

\begin{table*}[h]
\scriptsize
\centering
\caption{\textbf{MSC (\%) and MSV (\%) after four image transformations.}}
\begin{adjustbox}{max width=1.0\linewidth}
\begin{tabular}{c|c|cc|cc|cc|cc|cc}
 \Xhline{1pt}
\multirow{3}{*}{\textbf{\begin{tabular}[c]{@{}c@{}}GAN\\ Structure\end{tabular}}} & \multirow{3}{*}{\textbf{Method}} & \multicolumn{2}{c|}{\textbf{Target GAN}} & \multicolumn{8}{c}{\textbf{Image Transformation}} \\ \cline{3-12} 
 &  & \multirow{2}{*}{MSC $\Uparrow$} & \multirow{2}{*}{MSV $\Uparrow$} & \multicolumn{2}{c|}{Noise} & \multicolumn{2}{c|}{Blur} & \multicolumn{2}{c|}{Compression} & \multicolumn{2}{c}{Crop} \\ \cline{5-12} 
 &  &  &  & MSC $\Uparrow$ & MSV $\Uparrow$ & MSC $\Uparrow$ & MSV $\Uparrow$ & MSC $\Uparrow$ & MSV $\Uparrow$ & MSC $\Uparrow$ & MSV $\Uparrow$ \\ \hline
\multirow{5}{*}{StarGAN} & \texttt{AE-I} & 100.00 & 100.00 & 100.00 & 1.00 & 100.00 & 0.00 & 100.00 & 0.00 & 100.00 & 0.00 \\
 & \texttt{AE-D} & 100.00 & 100.00 & 100.00 & 100.00 & 100.00 & 100.00 & 100.00 & 100.00 & 3.00 & 100.00 \\
 & \texttt{CFP-AE} & 100.00 & 100.00 & 67.20 & 77.60 & 83.20 & 84.60 & 89.20 & 89.00 & 80.00 & 81.60 \\
 & \texttt{CFP-iBDv1} & 95.52 & 94.12 & 94.75 & 93.12 & 94.85 & 93.57 & 95.03 & 93.75 & 95.03 & 93.57 \\
 & \texttt{CFP-iBDv2} & 92.87 & 90.05 & 92.83 & 86.02 & 92.05 & 87.57 & 92.30 & 89.70 & 92.25 & 90.62 \\ \hline
\multirow{5}{*}{AttGAN} & \texttt{AE-I} & 100.00 & 100.00 & 100.00 & 1.00 & 100.00 & 0.00 & 100.00 & 0.00 & 100.00 & 0.00 \\
 & \texttt{AE-D} & 100.00 & 14.00 & 100.00 & 18.00 & 100.00 & 14.00 & 100.00 & 14.00 & 4.00 & 97.00 \\
 & \texttt{CFP-AE} & 100.00 & 100.00 & 67.20 & 82.00 & 80.20 & 73.00 & 86.80 & 84.80 & 67.60 & 76.80 \\
 & \texttt{CFP-iBDv1} & 93.40 & 92.45 & 92.60 & 91.20 & 92.75 & 92.07 & 92.90 & 92.52 & 93.33 & 91.57 \\
 & \texttt{CFP-iBDv2} & 91.03 & 90.70 & 91.38 & 80.32 & 91.40 & 84.10 & 91.58 & 89.30 & 91.58 & 88.82 \\ \hline
\multirow{5}{*}{STGAN} & \texttt{AE-I} & 98.00 & 100.00 & 100.00 & 0.00 & 100.00 & 0.00 & 100.00 & 0.00 & 100.00 & 0.00 \\
 & \texttt{AE-D} & 100.00 & 34.00 & 100.00 & 41.00 & 100.00 & 39.00 & 100.00 & 36.00 & 2.00 & 100.00 \\
 & \texttt{CFP-AE} & 100.00 & 100.00 & 85.50 & 77.00 & 94.80 & 41.20 & 96.00 & 65.20 & 84.20 & 57.00 \\
 & \texttt{CFP-iBDv1} & 93.53 & 91.57 & 92.53 & 89.00 & 92.98 & 90.72 & 93.30 & 91.45 & 93.48 & 90.95 \\
 & \texttt{CFP-iBDv2} & 92.20 & 90.18 & 91.53 & 86.40 & 91.48 & 86.55 & 91.75 & 88.15 & 91.58 & 88.80 \\
 \Xhline{1pt}
\end{tabular}
\end{adjustbox}
\label{tab:imagetrans}
\vspace{-5pt}
\end{table*}

\noindent\textbf{Persistency against model transformations.} We assume the adversary can have access to the corresponding discriminator of the stolen generator to facilitate the following experiments. However, it is not realistic, because the discriminator will be discarded after the training process. Therefore, the adversary we consider in this section is very strong. We apply pruning and fine-tuning\footnote{Fine-tuning GANs is actually not practical for an adversary to perform, as it requires the discriminator, which is kept secret by the model owner (see Appendix~\ref{appendix:power}). To demonstrate the strong persistency of our method, we still evaluate this impractical attack.} to moderately alter the GAN model. We also tried model quantification, which could significantly decrease the model usability~\cite{ganquan} (see Appendix~\ref{appendix:pruning}). So we ignore such operation. (1) For model fine-tuning, we refine the model with different epochs (10, 20 and 30) using the same training set\footnote{Fine-tuning a GAN using a different dataset of the same distribution will give the same conclusion. Fine-tuning using a dataset of different distributions is a challenging task in computer vision, and there are no satisfactory methods for us to follow.}. Such a setting is commonly used in previous works, and also in line with the adversary's capability in this paper. The learning rate is different for fine-tuning different model structures to avoid the collapse: $9.99e^{-5}$ for StarGAN, $1e^{-4}$ for AttGAN, $2e^{-5}$ for STGAN, which all follow the learning rate adjustment in the original papers. (2) For model pruning, we consider two compression ratios (0.2 and 0.4). Experiments show that a compression ratio higher than 0.4 can cause significant accuracy degradation for GAN models (see Appendix~\ref{appendix:pruning}).

As shown in Table~\ref{tab:robust}, \texttt{AE-I} can hardly resist these transformations because the above attacks fundamentally change the generation details of the target model, while the effectiveness of \texttt{AE-I} highly depends on the invariance of these details. \texttt{AE-D} will benefit from these model operations, which can further distort the model output and decrease the PSNR value. But this is still not enough for verifying AttGAN and STGAN. In contrast, our methods achieve satisfactory persistency under these modifications.

\noindent\textbf{Persistency against image transformations.} We evaluate the impact of image transformations. We first tried to transform the model input, which significantly degrades the quality of output images and is impractical for the adversary (see Appendix~\ref{appendix:pruning}). So we mainly consider the transformation of model output. We adopt four popular operations: \textit{adding Gaussian noises} (with mean $\mu=0$ and standard deviation $\sigma=0.1$), \textit{Gaussian blurring} (with a kernel size of 5), \textit{JPEG compression} (with a compression ratio of 35\%), and \textit{center cropping} (from $128\times128$ to $100\times100$). These transformations will still maintain the quality of the images. Table~\ref{tab:imagetrans} reports the Match scores. We observe \texttt{AE-I} is not robust at all, as these operations can significantly compromise the details of the images and invalidate the verification process. For our approach, \texttt{CFP-AE} is less effective for STGAN because the output of STGAN is more sensitive than other models, due to its adaptive selection structure giving more details in the output. In contrast, \Nameb and \Name perform the best, as the backdoor classifier together with the invisible backdoor samples are more robust against these operations, further enhanced by the data augmentation during fingerprint embedding. We also measure the impacts of different transformation strengths and other types of transformation operations in Appendix~\ref{appendix:pruning}, which has similar conclusions.

\vspace{-5pt}
\subsection{Stealthiness Analysis}
\label{sec:stealth-analysis}
\vspace{-5pt}

\setcounter{figure}{5}
\begin{figure*}[ht]
\vspace{-15pt}
\centering
\includegraphics[width=\linewidth]{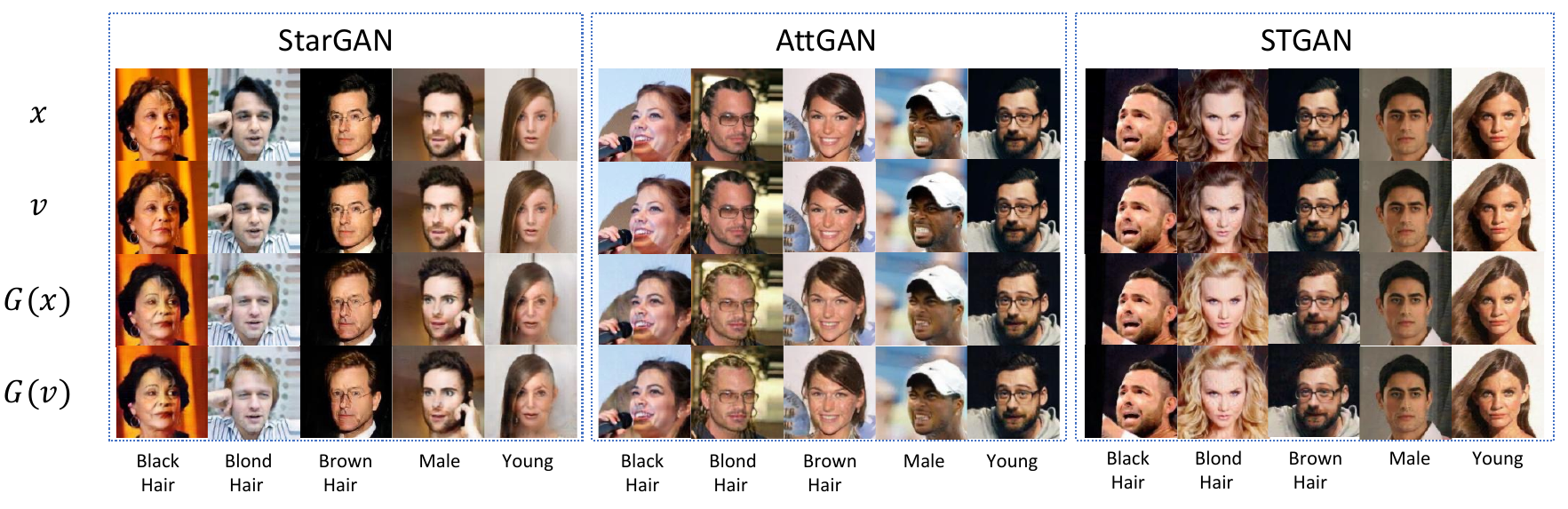}
\vspace{-15pt}
\caption{\textbf{Fingerprint visualization.} (a) Clean sample $x$; (b) Verification sample $v$; (c) GAN output of clean sample $G(x)$; (d) GAN output of verification sample $G(v)$. If the input agrees with the modifying attribute's label, the model will flip this label and modify the input with the flipped label.}
\label{fig:compare_new_title}
\vspace{-8pt}
\end{figure*}

We assess the stealthiness property from three perspectives. Note that our three methods share the same verification samples, as they use the same $\mathbf{F_{gen}}$. So we use \texttt{CFP-*} to denote any of our methods. We provide results on attribute editing tasks.

\vspace{3pt}
\noindent\textbf{Sample space indistinguishability.} 
Fig.~\ref{fig:compare_new_title} visually compares the verification query-response images with the ground-truth (normal images) for three different GANs using our proposed \texttt{CFP-*}. Visualizations of \texttt{CFP-*} for other GANs, and \texttt{AE-I} and \texttt{AE-D} can be found in Appendix~\ref{appendix:vs}. We observe that the perturbations added to the verification samples and model output samples are imperceptible. This confirms the effectiveness of the objective function in Equation~\ref{equ:loss1}. 

Quantitatively, Table~\ref{tab:dist_detail} shows the peak signal-to-noise ratio (PSNR) and structural similarity (SSIM)~\cite{psnrssim} between the input of clean and verification samples, as well as between their output samples. 
According to~\cite{LinhNA20,HongF0F20}, two pictures with PSNR$>35$ or SSIM$>0.95$ can be considered as the same in human vision. We observe that although \texttt{AE-I} and \texttt{AE-D} have indistinguishability for the input samples, their output images are significantly different from ground-truth ones. In contrast, our verification samples meet the visual indistinguishability from normal samples for both model inputs and outputs. This improves the concealment of ownership verification and makes it difficult for adversaries to distinguish verification samples from visual changes.

\begin{table*}[htbp]
\scriptsize
\centering
\caption{\textbf{PSNR and SSIM of the verification and clean input ($v$, $x$) and output ($G(v)$, $G(x)$) images for different edited attributes}. (``-'' in \texttt{AE-D} indicates we are not able to find the qualified verification samples with $\mathbf{F_{gen}}$.)}
\label{tab:dist}
\begin{adjustbox}{max width=1.\linewidth}
\begin{tabular}{c|c|ccccc|ccccc|ccccc}
\Xhline{1pt}
\multicolumn{2}{c|}{\multirow{2}{*}{\textbf{Similarity}}} & \multicolumn{5}{c|}{\textbf{StarGAN}} & \multicolumn{5}{c|}{\textbf{AttGAN}} & \multicolumn{5}{c}{\textbf{STGAN}} \\ \cline{3-17} 
\multicolumn{2}{c|}{} & A1 & A2 & A3 & A4 & A5 & A1 & A2 & A3 & A4 & A5 & A1 & A2 & A3 & A4 & A5 \\ \Xhline{1pt}
\multirow{3}{*}{PSNR$(v,x)$} & \texttt{AE-I} & 43.84 & 43.73 & 43.64 & 43.72 & 43.99 & 39.92 & 38.68 & 38.65 & 39.60 & 40.07 & 38.27 & 38.23 & 39.50 & 39.73 & 39.36 \\  
 & \texttt{AE-D} & 33.62 & 33.67 & 33.68 & 33.57 & - & 33.70 & 33.72 & 33.62 & 34.09 & 34.24 & 33.35 & 33.81 & 33.67 & - & 33.39 \\ 
 & \texttt{CFP-*} & 41.54 & 42.38 & 42.34 & 41.12 & 40.86 & 47.50 & 45.54 & 46.41 & 46.16 & 46.41 & 46.22 & 44.08 & 43.48 & 44.56 & 44.64 \\ \hline
\multirow{3}{*}{SSIM$(v,x)$} & \texttt{AE-I} & 0.99 & 0.99 & 0.99 & 0.99 & 0.99 & 0.96 & 0.95 & 0.95 & 0.96 & 0.96 & 0.97 & 0.97 & 0.98 & 0.97 & 0.98 \\
 & \texttt{AE-D} & 0.89 & 0.89 & 0.89 & 0.89 & - & 0.90 & 0.90 & 0.90 & 0.91 & 0.90 & 0.90 & 0.90 & 0.91 & - & 0.94 \\ 
 & \texttt{CFP-*} & 0.98 & 0.98 & 0.98 & 0.98 & 0.98 & 0.99 & 0.99 & 0.99 & 0.99 & 0.99 & 0.99 & 0.99 & 0.99 & 0.99 & 0.99 \\ \hline
\multirow{3}{*}{PSNR$(G(v),G(x))$} & \texttt{AE-I} & 22.44 & 18.28 & 23.37 & 24.59 & 23.98 & 30.83 & 27.79 & 29.04 & 29.43 & 29.66 & 31.78 & 33.82 & 36.32 & 37.29 & 36.87 \\ 
 & \texttt{AE-D} & 10.99 & 10.07 & 10.85 & 10.77 & - & 23.94 & 22.92 & 24.51 & 25.65 & 28.29 & 22.50 & 20.12 & 25.71 & - & 29.62 \\ 
 & \texttt{CFP-*} & 37.75 & 38.00 & 38.12 & 37.37 & 37.20 & 45.33 & 43.33 & 44.38 & 44.46 & 44.36 & 44.53 & 42.78 & 42.81 & 43.94 & 44.01 \\ \hline
\multirow{3}{*}{SSIM$(G(v),G(x))$} & \texttt{AE-I} & 0.84 & 0.79 & 0.87 & 0.87 & 0.85 & 0.95 & 0.92 & 0.92 & 0.94 & 0.94 & 0.96 & 0.97 & 0.98 & 0.98 & 0.98 \\
 & \texttt{AE-D} & 0.40 & 0.38 & 0.41 & 0.40 & - & 0.86 & 0.85 & 0.87 & 0.89 & 0.90 & 0.85 & 0.84 & 0.90 & - & 0.93 \\ 
 & \texttt{CFP-*} & 0.97 & 0.97 & 0.97 & 0.96 & 0.96 & 0.99 & 0.99 & 0.99 & 0.99 & 0.99 & 0.99 & 0.99 & 0.99 & 0.99 & 0.99 \\
 \Xhline{1pt}
\end{tabular}
\end{adjustbox}
\label{tab:dist_detail}
\vspace{-10pt}
\end{table*}

\begin{table}[ht]
\centering
\caption{\textbf{The cumulative probability of standard deviation of feature maps for different types of samples.}}
\begin{adjustbox}{max width=1.0\linewidth}
\begin{tabular}{c|cccccccc}
\Xhline{1pt}
\multirow{2}{*}{\textbf{Method}} & \multicolumn{8}{c}{Standard deviation of feature maps} \\ \cline{2-9} 
 & 0.650 & 0.675 & 0.700 & 0.725 & 0.750 & 0.775 & 0.800 & 0.825 \\ \Xhline{1pt}
Normal & 0\% & 3\% & 9\% & 15\% & 38\% & 63\% & 92\% & 100\% \\
\texttt{AE-I} & 0\% & 2\% & 19\% & 58\% & 97\% & 100\% & 100\% & 100\% \\ 
\texttt{AE-D} & 0\% & 18\% & 64\% & 99\% & 100\% & 100\% & 100\% & 100\% \\
\texttt{CFP-*} & 0\% & 3\% & 14\% & 37\% & 69\% & 86\% & 98\% & 100\%\\
\Xhline{1pt}
\end{tabular}
\end{adjustbox}
\label{tab:diff}
\vspace{-8pt}
\end{table}

\begin{table}[ht]
\centering
\caption{\textbf{AUC for detecting verification samples.}}
\begin{adjustbox}{max width=1.0\linewidth}
\begin{tabular}{c|ccc|ccc}
\Xhline{1pt}
\multirow{2}{*}{\textbf{Method}} & \multicolumn{3}{c|}{\textbf{Scenario 1}} & \multicolumn{3}{c}{\textbf{Scenario 2}} \\ \cline{2-7}
 & $l_2$ & $l_1$ & $l_\infty$ & $l_2$ & $l_1$ & $l_\infty$ \\ \hline
\texttt{AE-D} & 85.74 & 99.82 & 86.39 & 55.46 & 65.78 & 55.85 \\ \hline
\texttt{AE-I} & 75.96 & 80.37 & - & 48.19 & 52.69 & -  \\ \hline
\texttt{CFP-*} & 59.00 & - & - & 51.51 & - & - \\  \Xhline{1pt}
\end{tabular}
\end{adjustbox}
\label{tab:auc}
\vspace{-13pt}
\end{table}

\vspace{3pt}
\noindent\textbf{Feature space indistinguishability.} An adversary may try to monitor the intermediate features of the inference process to detect the verification samples. Anomalous samples usually result in unique behaviors in the feature space, which has been exploited to detect adversarial attacks \cite{activation_spaces,lee_simple_2018,model_mutation}. Specifically, we produce 100 samples for each category (normal, \texttt{AE-I}, \texttt{AE-D}, and \texttt{CFP-*}). We compute the standard deviation of features for each sample, and then the cumulative probabilities among these 100 samples in one category. If the cumulative probability distribution of one category is closer to that of normal samples, it is harder for the adversary to identify such verification samples in the feature space. Table~\ref{tab:diff} shows the statistical results. 
The stealthiness of the verification samples from \texttt{CFP-*} is much better than that of \texttt{AE-I} and \texttt{AE-D} since its distribution is closer to the normal one's.  
This is because the construction of our verification samples is forced to minimize the distance $\Vert G_{v,j}, G_{x,j} \Vert$, which fundamentally ensures the consistency of the distribution in the feature space between verification and normal samples.

\vspace{3pt}
\noindent\textbf{Indistinguishability from machine learning detectors.}
An adversary can employ an anomaly detector to identify verification samples. To show this feasibility, we introduce two detecting scenarios for an adversary. \textbf{Scenario 1}: The detection model accepts differences between inference samples and the corresponding results of the GAN model as inputs. \textbf{Scenario 2}: The detection model accepts inference samples as inputs. We train lightweight one-class models based on HRN~\cite{hrn} only from clean samples for each scenario, respectively, since the adversary does not have information on the verification samples. The verification samples are secretly stored in the trusted third party. The adversary can only adopt clean data to train a detector to detect verification samples. Specifically, the prediction of the detector is projected to a non-linear space $(0,1)$ through a function $\mathrm{sigmoid}(x)=1/{(1+\exp^{-x})}$. The detection result of normal samples is expected to be close to 1, while that of verification samples is expected to be close to 0. The detector in our experiment is ResNet34~\cite{he_deep_2016} trained with the one-class detection method~\cite{hrn}.
Table~\ref{tab:auc} shows the performance of our anomaly detector under two scenarios. We adopt the AUC (Area Under the ROC Curve) metric: a higher AUC score means higher detection accuracy, while a score of 50 indicates the detector has similar performance as random guess. We observe the it is challenging to perform detection just based on the inference samples (Scenario 2) for all the methods. However, if the adversary adopts the difference between the inference samples and their corresponding outputs (Scenario 1), the anomaly detector can identify the verification samples based on \texttt{AE-D} and \texttt{AE-I} with very high accuracy, while the detection accuracy of \texttt{CFP-*} is still low. This demonstrates that our method is a stealthy approach for IP protection of GAN models. 

\vspace{-5pt}
\subsection{Generalize to Other Tasks}
\vspace{-5pt}

We evaluate \texttt{CFP-*} on the domain translation task and super resolution task to prove that our scheme is general to various I2I GANs. The results are shown in Tables~\ref{tab:u_two} and~\ref{tab:r_two}. We observe a strong consistency among different I2I tasks. Our methods can successfully verify protected GANs and discriminate unrelated GANs. They are robust against various model modifications\footnote{We do not consider image transformations because these transformations significantly change the outputs under these two I2I tasks, violating our threat model.} as well. Therefore, our methods are general for different I2I GANs.

\begin{table}[t]
\scriptsize
\centering
\caption{\textbf{MSC (\%) and MSV (\%) for verifying different GANs. The target GANs are C1 and SRResNet for domain translation and super resolution, respectively. C2 and C3 are for domain translation. RRDBNet and EDSR are for super resolution.}}
\begin{adjustbox}{max width=0.95\linewidth}
\begin{tabular}{c|c|cc|cc}
 \Xhline{1pt}
\multirow{3}{*}{\textbf{Task}} & \multirow{3}{*}{\textbf{Method}} & \multicolumn{2}{c|}{\textbf{Target GAN}} & \multicolumn{2}{c}{\textbf{Non-target GAN}} \\ \cline{3-6} 
 &  & \multirow{2}{*}{MSC $\Uparrow$} & \multirow{2}{*}{MSV $\Uparrow$} & \multicolumn{1}{c|}{C2 / RRDBNet} & C3 / EDSR \\ \cline{5-6} 
 &  &  &  & \multicolumn{1}{c|}{MSV $\Downarrow$} & MSV $\Downarrow$ \\ \hline
\multirow{3}{*}{\begin{tabular}[c]{@{}c@{}}Domain\\ Translation\end{tabular}} & \texttt{CFP-AE} & 100 & 88 & \multicolumn{1}{c|}{51} & 44 \\
 & \texttt{CFP-iBDv1} & 93 & 97 & \multicolumn{1}{c|}{41} & 30 \\
 & \texttt{CFP-iBDv2} & 99 & 97 & \multicolumn{1}{c|}{38} & 27 \\ \hline
\multirow{3}{*}{\begin{tabular}[c]{@{}c@{}}Super\\ Resolution\end{tabular}} & \texttt{CFP-AE} & 100 & 100 & \multicolumn{1}{c|}{56} & 45 \\
 & \texttt{CFP-iBDv1} & 100 & 100 & \multicolumn{1}{c|}{73} & 49 \\
 & \texttt{CFP-iBDv2} & 100 & 100 & \multicolumn{1}{c|}{48} & 35 \\
 \Xhline{1pt}
\end{tabular}
\end{adjustbox}
\label{tab:u_two}
\vspace{-10pt}
\end{table}

\begin{table*}[t]
\scriptsize
\centering
\caption{\textbf{MSC (\%) and MSV (\%) after two model transformations. The target GANs are C1 and SRResNet for domain translation and super resolution, respectively.}}
\begin{adjustbox}{max width=0.95\linewidth}
\begin{tabular}{c|c|cc|cccccc|cccc}
 \Xhline{1pt}
\multirow{3}{*}{\textbf{Task}} & \multirow{3}{*}{\textbf{Method}} & \multicolumn{2}{c|}{\multirow{2}{*}{\textbf{Target GAN}}} & \multicolumn{6}{c|}{\textbf{Fine-tuning (epochs)}} & \multicolumn{4}{c}{\textbf{Pruning (compression ratio)}} \\ \cline{5-14} 
 &  & \multicolumn{2}{c|}{} & \multicolumn{2}{c|}{10} & \multicolumn{2}{c|}{20} & \multicolumn{2}{c|}{30} & \multicolumn{2}{c|}{0.2} & \multicolumn{2}{c}{0.4} \\ \cline{3-14} 
 &  & MSC $\Uparrow$ & MSV $\Uparrow$ & MSC $\Uparrow$ & \multicolumn{1}{c|}{MSV $\Uparrow$} & MSC $\Uparrow$ & \multicolumn{1}{c|}{MSV $\Uparrow$} & MSC $\Uparrow$ & MSV $\Uparrow$ & MSC $\Uparrow$ & \multicolumn{1}{c|}{MSV $\Uparrow$} & MSC $\Uparrow$ & MSV $\Uparrow$ \\ \hline
\multirow{3}{*}{\begin{tabular}[c]{@{}c@{}}Domain\\ Translation\end{tabular}} & \texttt{CFP-AE} & 100 & 88 & 97 & \multicolumn{1}{c|}{88} & 98 & \multicolumn{1}{c|}{88} & 91 & 88 & 96 & \multicolumn{1}{c|}{88} & 83 & 88 \\
 & \texttt{CFP-iBDv1} & 93 & 97 & 91 & \multicolumn{1}{c|}{98} & 95 & \multicolumn{1}{c|}{94} & 94 & 91 & 96 & \multicolumn{1}{c|}{92} & 95 & 87 \\
 & \texttt{CFP-iBDv2} & 99 & 97 & 94 & \multicolumn{1}{c|}{95} & 96 & \multicolumn{1}{c|}{91} & 95 & 85 & 100 & \multicolumn{1}{c|}{90} & 97 & 85 \\ \hline
\multirow{3}{*}{\begin{tabular}[c]{@{}c@{}}Super\\ Resolution\end{tabular}} & \texttt{CFP-AE} & 100 & 100 & 91 & \multicolumn{1}{c|}{100} & 93 & \multicolumn{1}{c|}{100} & 89 & 100 & 52 & \multicolumn{1}{c|}{100} & 39 & 92 \\
 & \texttt{CFP-iBDv1} & 100 & 100 & 100 & \multicolumn{1}{c|}{100} & 100 & \multicolumn{1}{c|}{100} & 100 & 100 & 100 & \multicolumn{1}{c|}{100} & 100 & 99 \\
 & \texttt{CFP-iBDv2} & 100 & 100 & 100 & \multicolumn{1}{c|}{100} & 100 & \multicolumn{1}{c|}{100} & 100 & 100 & 100 & \multicolumn{1}{c|}{100} & 100 & 97 \\
 \Xhline{1pt}
\end{tabular}
\end{adjustbox}
\label{tab:r_two}
\vspace{-5pt}
\end{table*}

\vspace{-5pt}
\subsection{Summary}
\vspace{-5pt}

Table~\ref{tab:summary} summarizes the comparisons of those methods from the above evaluations. There are five levels to assess each property of each method. \texttt{AE-I} is effective for fingerprinting the GAN model, but not robust enough against model pruning, fine-tuning or image transformations. \texttt{AE-D} cannot guarantee the high quality of verification samples on other GAN models, leading to low MSV scores. In Appendix~\ref{appendix:vs}, we show the outputs of verification samples for three GANs, which reveal that
\texttt{AE-D} is not a stable and general fingerprinting method. Besides, \texttt{AE-I} and \texttt{AE-D} are not stealthy, which gives an adversary more chances to detect the verification samples and manipulate the results. For our proposed scheme, \texttt{CFP-AE} is not good at resisting image transformations. With the introduction of the invisible backdoor technique for fingerprint embedding, \texttt{CFP-iBDv1} and \texttt{CFP-iBDv2} can significantly improve the effectiveness and persistency. The two novel loss function terms in \Name can further increase the distinguisability between target and non-target GAN models. The three methods also give much better stealthiness in both the sample space and feature space.

\begin{table}[h]
\centering
\caption{\textbf{Assessment summary of each method.} (Excellent $>$ Good $>$ Fair $>$ Poor $>$ Bad)}
\begin{adjustbox}{max width=1.0\linewidth}
\begin{tabular}{c|c|cc|ccc}
 \Xhline{1pt}
\multirow{2}{*}{\textbf{Method}} & \multirow{2}{*}{\textbf{Distinctness}} & \multicolumn{2}{c|}{\textbf{Persistency}} & \multicolumn{3}{c}{\textbf{Stealthiness}} \\ \cline{3-7} 
 & & Model trans. & Image trans. & Sample space & Feature space  & Detection \\ \hline
\texttt{AE-I} & Excellent & Bad & Bad & Fair & Bad & Bad \\
\texttt{AE-D}& Poor & Fair & Fair  & Poor & Bad & Bad\\
\texttt{CFP-AE} & Poor & Excellent & Good & Excellent & Good & Good\\
\texttt{CFP-iBDv1} & Fair & Excellent & Excellent & Excellent & Good & Good\\
\texttt{CFP-iBDv2} & Good & Excellent & Excellent & Excellent & Good & Good\\
 \Xhline{1pt}
\end{tabular}
\end{adjustbox}
\label{tab:summary}
\vspace{-10pt}
\end{table}

\vspace{-5pt}
\section{Discussions}
\vspace{-5pt}

\subsection{Limitations and Future Work}
\vspace{-5pt}


\noindent\textbf{Fingerprinting other types of generative models.}
This paper mainly focuses on the protection of I2I GANs. There are other types of GANs, e.g., noise-to-image translation, models for synthesizing audios, texts, etc. We expect our scheme to be general and extensible for those models as well. We will consider this as future work. On the other hand, recent diffusion models~\cite{ho_denoising_2020} are proposed as a more advanced generative model. However, the forward process requires many sampling steps to obtain high-quality images, which makes the optimization process during fingerprint generation computationally impossible. Besides, the diffusion model can be sampled randomly to generate various outputs, making the verification samples invalid. Therefore, our scheme does not fit diffusion models. A new solution dedicated to diffusion models is desired as future work.

\noindent\textbf{Protection against model extraction attacks.}
Although a few works about IP protection of classification models \cite{lukas2019deep,jia2021entangled,peng2022fingerprinting} evaluate model extraction attacks, they are not included in our threat model. The main reason is that extracting a GAN model requires the adversary to have much more significant amounts of computing resources than stealing a classification model, and is much easier to defeat by simply adding small scales of Gaussian noise to the output \cite{hu2021stealing}. Besides, \cite{hu2021stealing} only presents the attacks against noise-to-image GAN models, while the feasibility of extracting I2I models is unknown. How to design more resource-efficient model extraction attacks and evaluate the effectiveness of our scheme against them are interesting future directions.

\noindent\textbf{Alternative schemes.}
We evaluate existing adversarial attacks for GAN models \cite{aed3,aed2,aed,aes2,aes} as the fingerprint baselines, and show their limitations in stealthiness and persistency. An alternative direction is to seek for more robust and stealthy attacks for fingerprinting GANs. Adversarial attacks against GANs are much less studied, and we could not find a satisfactory solution. On the other hand, the intrinsic fingerprint of GANs in the frequency domain is not robust against the changing brightness of the image. Therefore, such a fingerprint cannot be used in I2I GANs, such as super-resolution, denoising, and colorizing. We urge researchers to explore this direction for both effective attacks and fingerprinting solutions. Nevertheless, our novel scheme provides a different perspective with off-the-shelf methodologies.

\vspace{-5pt}
\subsection{Related Works}
\label{sec:related}
\vspace{-5pt}

We discuss some relevant works and highlight their differences from our solution. 

\vspace{3pt}
\noindent\textbf{Watermarking GANs.}
Compared to classification models, IP protection of GAN models is much less explored. Prior works \cite{ong2021protecting,fei_supervised_2022,lin_cycleganwm_2022} designed watermarking solutions for GAN models. To embed a watermark into a protected GAN, the model owner needs to train the model from scratch, which is less practical for an already trained GAN. As discussed in Section~\ref{sec:introduction}, watermarking has the limitations of usability and applicability~\cite{cao2019ipguard,peng2022fingerprinting}, which can be solved by fingerprinting.

\vspace{3pt}
\noindent\textbf{Watermarking Diffusion Models.}
There are several recent works~\cite{liu_watermarking_2023,wen_tree-ring_2023,fernandez_stable_2023} focusing on the IP protection of diffusion models~\cite{dm}. Diffusion models can synthesize high-quality images from noise or text descriptions, or perform I2I translation. There are more and more applications based on diffusion models. These methods are mainly based on backdoor techniques, making the diffusion model generate samples containing specific patterns in the signal domain or pixel domain, which can be recognized by a detector. Embedding such backdoors require training or finetuning the diffusion model, which is costly in terms of time and resources. It is interesting to extend our solution to fingerprinting diffusion models.

\vspace{3pt}
\noindent\textbf{Using GANs for IP protection.}
Some works utilized GANs to enhance or defeat IP protection methods. For instance, GANs are used to generate watermarks for BERT language models \cite{AbdelnabiF21}, and identify and remove watermarks from classification models \cite{wang2021detect}. Different from those works, our solutions focus on protecting I2I GANs. 

\vspace{3pt}
\noindent\textbf{Meta learning for fingerprinting}.
Pan et al.~\cite{metav} propose a meta-learning-based fingerprinting scheme, which is a task-agnostic framework independent of the tasks. They adopt a number of positive and negative suspect models, where the positive suspect models are derived from the protected model based on pruning, fine-tuning, and distillation, and the negative models are models different from the target model for different training data or model structures. Although their framework shows robustness against various attacks, we argue that it is not practical to protect GANs for commercial use. In Section~\ref{sec:time}, we show the time to train a high-quality GAN. The heavy time cost makes this framework unrealistic in protecting state-of-the-art GANs.

\vspace{3pt}
\noindent\textbf{Detecting and attributing GAN-generated images.}
Some works~\cite{yu2019attributing,yu2020artificial,yu2020responsible} leveraged fingerprints to detect GAN-generated images and trace their sources.
However, they are not quite applicable to fingerprint GAN models for IP protection. For instance, \cite{yu2020artificial,yu2020responsible} require the model owner to modify the GAN model training process (e.g., training loss and training data) to have the capability of embedding fingerprints in the output images, which violates the requirement of model fingerprinting. In \cite{yu2019attributing}, 
the fingerprint in the output image is very sensitive to model transformations: ``Even GAN training sets that differ in just one image can lead to distinct GAN instances \cite{yu2019attributing}.'' As a result, an adversary can just use a different training set to fine-tune the target GAN model to invalidate the fingerprint. In contrast, our methods do not need to modify the model and exhibit higher unremovability. \cite{nie2023attributing} adds a fingerprint into the generative model by modifying the model structure, which is obvious for an adversary to find out such a modification. This method also requires the generative model to use the latent space during the generation process, which is not general for all GANs.

\vspace{-7pt}
\section{Conclusion}
\vspace{-7pt}

We propose a novel scheme to fingerprint GAN models for IP protection. We introduce a classifier to construct a composite model with the protected GAN. From this composite model, we craft verification samples as the fingerprint, and embed it in the classifier. The classifier can distinguish the target and non-target models in a stealthy and robust manner. We design three fingerprinting methodologies based on generative adversarial examples and invisible backdoor attacks. Extensive evaluations validate the effectiveness of our designs.

\vspace{-7pt}
\section{Acknowledgement}
\vspace{-7pt}

This study is supported under the RIE2020 Industry Alignment Fund – Industry Collaboration Projects (IAF-ICP) Funding Initiative, as well as cash and in-kind contribution from the industry partner(s).

\bibliographystyle{plain}
\bibliography{gan}

\appendices

\setcounter{table}{0}
\setcounter{figure}{0}
\setcounter{section}{0}



\section{Symbols and Remarks}\label{appendix:sr}

\renewcommand{\arraystretch}{1}
\begin{table}[h]
\centering
\caption{Some important symbols and their remarks.}
\label{tab:symbols}
\begin{adjustbox}{max width=1.0\linewidth}
\begin{tabular}{ll}
\hline
\textbf{Symbol} & \textbf{Remarks} \\ \hline
$mk$ & a secret marking key \\
$vk$ & a public verification key \\
$v^{(i)}$ & a verification sample \\
$x^{(i)}$ & a clean sample \\
$ts$ & a timestamp \\
$D$ & a sample space \\
$\bar{D}$ & a defined sample space \\
$L$ & a label space \\
$V$, $V'$ & verification sample sets \\
$V_L$, $V_L'$ & verification label sets \\
$F$ & a protected deep learning model \\
$F^s$ & a suspicious model, whether it is stolen or not \\
$G$ & the Generator of protected GAN \\
$G_{v^{(i)},j}$, $G_{x^{(i)},j}$ & the $j$-th feature maps in $G$ \\
$G(x)$ & the accurate model outputs \\
$G(x)^p$ & the perturbed model outputs by the adversary \\
$\mathcal{G}(x)$ & a set contains outputs of $G(x)$ and $G(x)^p$ \\
$f^{*}$ & a ground-truth classifier projecting $D$ to $L$ \\
$f$ & a normal classifier trained with $\mathcal{O}^{f^{*}}$ \\
$\hat{f}$ & $f$ after fingerprinted \\
$M$ & a composite deep learning model with $f$ and ${G}$ \\
$\hat{M}$ & a fingerprinted composite deep learning model with $\hat{f}$ and ${G}$ \\
$\mathcal{O}^{f^{*}}$ & an oracle truly answering each call to $f^{*}$ \\
$\mathcal{A}$, $\mathcal{T}$, $\mathcal{S}$ &  PPT algorithms \\ \hline
\end{tabular}
\end{adjustbox}
\vspace{-10pt}
\end{table}

\section{Adversary's Power}
\label{appendix:power}

Let us introduce the adversary's capability in practice with more details. Firstly, when the model owner provides the service to others to use the model, the owner only needs the Generator of the GAN. The Discriminator of the GAN is deprecated, and any user cannot have the access to it, as the Discriminator may be deleted after the GAN is trained. It means that the adversary cannot steal both Generator and Discriminator, and the adversary can only steal the Generator at most. Secondly, the adversary does not have the ability to train a GAN to obtain the same performance as the stolen one, otherwise the adversary has no need to steal other's model. It means that the adversary does not have the ability to restore the Discriminator from the Generator. When fine-tuning the Generator, the adversary must have the Generator and the corresponding Discriminator at the same time. So the fine-tuning process is not possible in practice. However, the adversary may steal the Discriminator by some adaptive ways, such as stealing the hard disk storing the Discriminator. That is why fine-tuning the Generator is an adaptive attack in this paper. Further, let us consider a situation where the adversary prunes the Generator first and fine-tunes it later. For a PPT adversary, it is impossible to fine-tune a pruned Generator, as the Discriminator is no longer aligned with the pruned Generator. If using an unaligned Discriminator, it is easy to cause collapse in fine-tuning and make the Generator give worse results. A PPT adversary cannot find a set of parameters for the Discriminator to cooperate with the new Generator, otherwise the adversary can remove the fingerprint in the Generator directly by finding another set of parameters for it. As it is impossible, we do not consider this method in our experiments.

\section{Proof of Theorem 1}
\label{sec:proofs}
(I) \textbf{Functionality-preserving}.  By the definition of the algorithm $\mathbf{F_{emb}}$, it outputs a model $\hat{M}$ that satisfies
\begin{small}
 \begin{equation*}
\begin{split}
&\mathop{Pr}\limits_{x\in \bar{D}\backslash V}[f^{*}(\mathcal{G}(x))\neq \mathbf{Classify}(\hat{M}, x)]\leq \epsilon, \mathrm{and}\\
&\mathop{Pr}\limits_{x\in V}[V_L(x)\neq \mathbf{Classify}(\hat{M}, x)]\leq \epsilon.
\end{split}
\end{equation*}
\end{small}
As a result, given  an error $\epsilon$, $\hat{M}$ classifies correctly for at least $(1-\epsilon)|\mathcal{V}|$ elements in $\mathcal{V}$, which is consistent with the argument that $\mathbf{Classify}$ outputs 1 if $\hat{M}$ disagrees with $\mathcal{V}$ on at most $\epsilon|\mathcal{V}|$ elements.

(II) \textbf{Unremovability}. As defined in Section~\ref{Strong Fingerprints}, we assume that no algorithms can generate an $\epsilon$-accurate model $N$ in the time $t$ of $f$,  where $t$ is much smaller than the time required to train a model with the same accuracy as $N$ using the algorithm $\mathbf{Train}$. In addition, we assume that the time taken by the adversary $\mathcal{A}$ to break the requirement of unremovability is approximately $t$. According to \textbf{Game 1}, $\mathcal{A}$  will output an $\epsilon$-accurate model  when it is given the knowledge of $\tilde {M}$ and $vk$, where  at least a $(1-\epsilon)$ fraction of the  elements in $V$ are classified correctly by  $\tilde {M}$. We first prove that the adversary's realization of this is independent of the key  $vk$. To achieve this, we construct a series of algorithms to gradually replace the verification samples in $vk$ with other random values. Specifically, consider the following algorithm $\mathcal{S}$:
\begin{itemize}
\begin{small}
\setlength\itemsep{0.1em}
\item [1.] Generate $M\leftarrow \mathbf{Train}(\mathcal{O}^{f^*}, \mathcal{G})$ and ($mk, vk$)$\leftarrow  \mathbf{KeyGen}()$.
\item  [2.] Compute $\hat{M}\leftarrow \mathbf{FP}(M, mk)$ and  run  $(\tilde{V},\tilde{V_L})=\tilde{\mathcal{V}}\leftarrow \mathbf{F_{gen}}(\mathcal{O}^{f^{*}}, G)$, where $\tilde{V}=\{\tilde{v}^{(i)}|i\in[n]\}$, $\tilde{V_L}=\{ \tilde{v_L}^{(i)}|i\in[n]\}$.
\item  [3.] Set $c_v^{(1)}\leftarrow \mathbf{Com}(\tilde{v}^{(1)}, h_v^{(1)})$,  $c_L^{(1)}\leftarrow \mathbf{Com}(\tilde{v_L}^{(1)}, h_L^{(1)})$, and $\tilde{vk}\leftarrow \{c_v^{(i)},  c_L^{(i)}\}_{i\in [n]}$. Then, compute  $\tilde{M}\leftarrow\mathcal{A}(\mathcal{O}^{f}, \tilde{vk}, \hat{M})$.
\end{small}
\end{itemize}

This algorithm replaces the first element in $vk$ with an independently generated random element, and then runs $\mathcal{A}$ on it.  Due to the statistical hiding  property of $\mathbf{Com}$, the output of $\mathcal{S}$ is statistically close to the output of $\mathcal{A}$ in the unremovablity experiment. Therefore,  we can further generate a series of hybrids $\mathcal{S}^{(2)}, \mathcal{S}^{(3)}\cdots, \mathcal{S}^{(n)}$ to change the 2nd to $n$-th elements in $vk$ in the same way. This means that the model $\tilde {M}$ generated by the adversary $\mathcal{A}$ must be independent of $vk$. Based on this, we consider the following algorithm $\mathcal{T}$:
\begin{itemize}
\begin{small}
\setlength\itemsep{0.1em}
\item  [1.] Compute $(mk, vk)\leftarrow \mathbf{KeyGen}()$.
\item [2.] Run the adversary and compute $\tilde{N}\leftarrow \mathcal{A}(\mathcal{O}^{f}, M, vk)$.
\end{small}
\end{itemize}

According to the above hybrid argument, the running time of the algorithm $\mathcal{T}$ is similar to that of $\mathcal{A}$, i.e., time $t$. Then it generates a model $\tilde{N}$ which does not contain the fingerprint. However,  this is contrary to the previous assumption about the persistence of strong fingerprints, i.e., $\mathcal{T}$  must also generate an $\epsilon$-accurate model given any model in the same time $t$.

(III) \textbf{Non-rewriteability}. Suppose there is a polynomial time algorithm $\mathcal{A}$ which can break the non-rewriteability requirement. This means that the timestamp $ts'$ owned by the adversary is generated earlier than the $ts$ of the model owner, and the model $\tilde{M}$ owned by the adversary also passes the trusted third party verification process. Obviously, if $\tilde{M}$ is built after $\hat{M}$, $ts'$ must be smaller than $ts$. This is because the trusted third party requires all model owners to upload the model as soon as possible after generating a complete composite model and use the upload time as the timestamp. The trusted third party will verify the copyright of the model and the legality of the timestamp. Therefore, it is impossible for the adversary to construct $ts'$ smaller than $ts$ without knowing the victim model $\hat{M}$, since the trusted third party needs to verify the copyright of the composite model bundled with $ts'$ while verifying the legitimacy of $ts'$.

\begin{figure*}[h]
\centering
\includegraphics[width=1.0\linewidth]{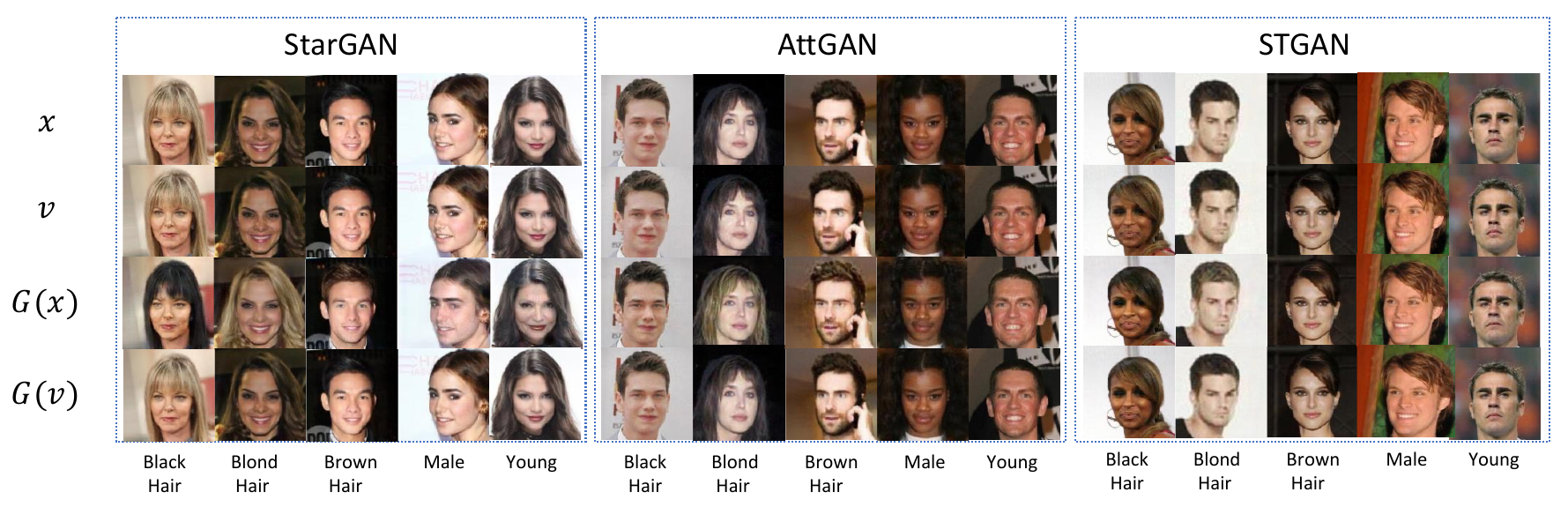}
\caption{Fingerprint visualization generated from \texttt{AE-I} for three attribute editing GANs with five edited attributes. (a) Clean sample $x$; (b) Verification sample $v$; (c) GAN output of clean sample $G(x)$; (d) GAN output of verification sample $G(v)$.}
\label{fig:ae}
\vspace{-10pt}
\end{figure*}

\begin{figure*}[h]
\centering
\includegraphics[width=1.0\linewidth]{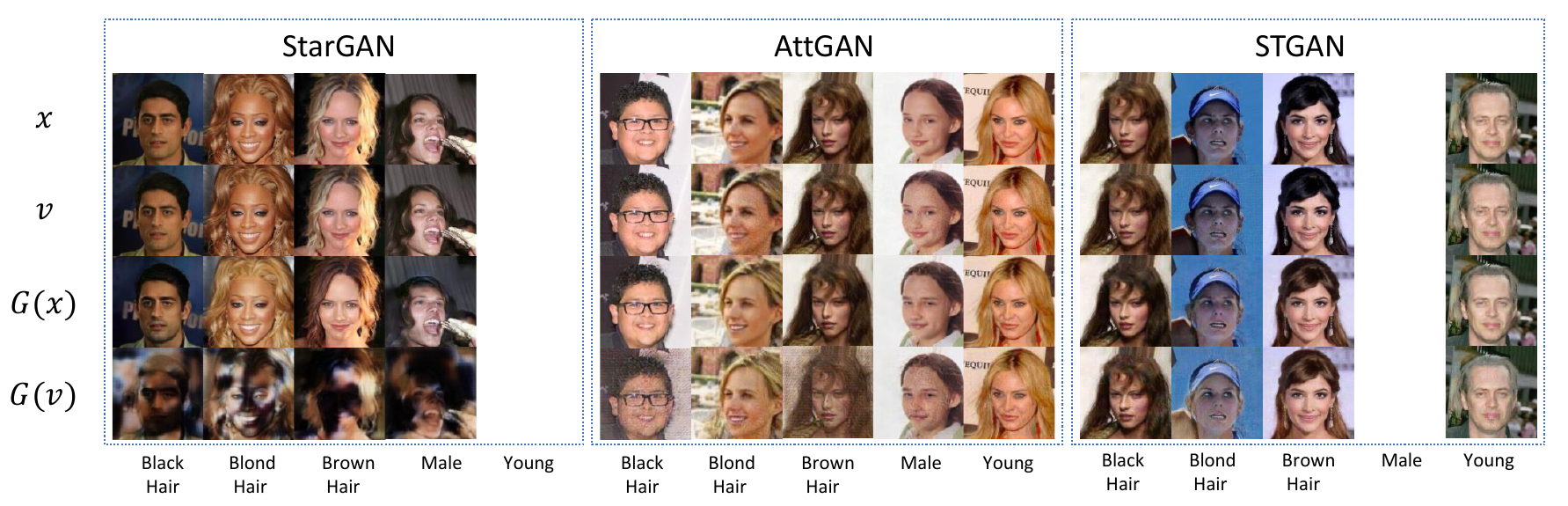}
\caption{Fingerprint visualization generated from \texttt{AE-D} for three attribute editing GANs with five edited attributes. Because there are no verification samples for some attributes, we leave these columns blank.}
\label{fig:ae-d}
\vspace{-15pt}
\end{figure*}

\begin{figure}[h]
\centering
\includegraphics[width=1.0\linewidth]{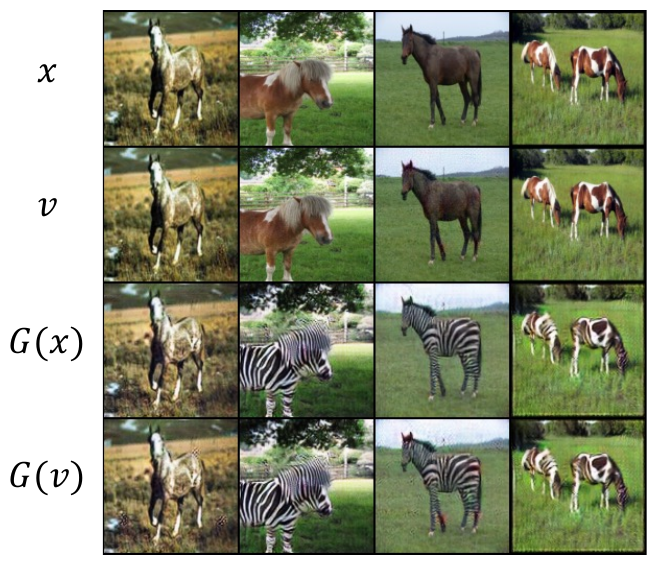}
\caption{Fingerprint visualization \texttt{CFP-*} for the domain translation GAN, C1.}
\label{fig:cyclegan}
\vspace{-10pt}
\end{figure}

\begin{figure}[h]
\centering
\includegraphics[width=1.0\linewidth]{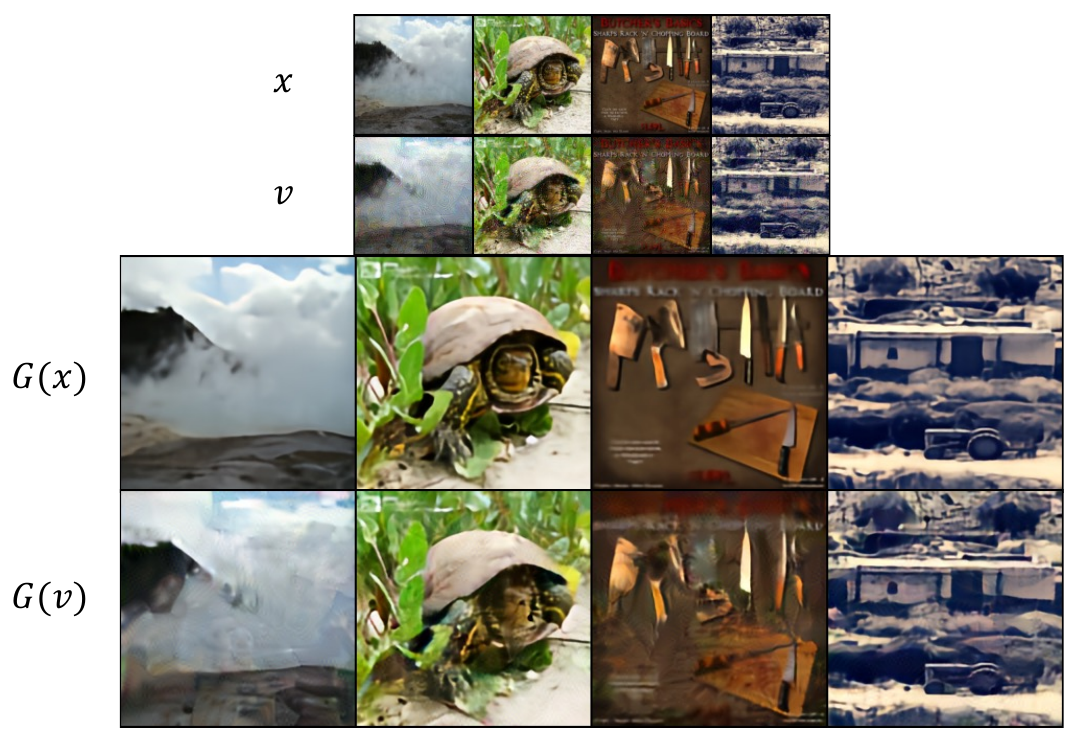}
\caption{Fingerprint visualization \texttt{CFP-*} for the super resolution GAN, SRResNet.}
\label{fig:sr}
\vspace{-10pt}
\end{figure}

\section{Verification Samples of Different Schemes}
\label{appendix:vs}

In this section, we show the verification samples of \texttt{AE-I} and \texttt{AE-D} in Fig.~\ref{fig:ae} and Fig.~\ref{fig:ae-d}, respectively. For \texttt{AE-I}, it is clear that all GANs can generate high quality outputs from the verification samples, and they are similar visually. However, \texttt{AE-D} does not have the same performance on AttGAN and STGAN as on StarGAN. Because AttGAN and STGAN have more stable generation structures, which means generating disrupted images by them are more difficult. On the other hand, \texttt{AE-D} still achieves a very high SSIM on AttGAN and STGAN, indicating it is not a stable and general fingerprinting scheme. In Fig.~\ref{fig:cyclegan}, and Fig.~\ref{fig:sr}, we show the visualization results of \texttt{CFP-*} for other tasks. The results prove the generalizability of our proposed fingerprinting schemes.

\section{High-resolution Facial Images of \texttt{CFP-*}}
In our experiments, we evaluate the performance of our proposed method by verifying whether the generated fingerprints satisfy the \textit{functionality-preserving}, \textit{unremovability}, and \textit{stealthiness} properties. Here, we present the high-resolution of GAN outputs of \texttt{CFP-*} in suffering various degradation, including model compression and corruptions with common image transformations.

\subsection {Verification Samples after Different GANs}
\begin{figure}[h]
\centering
\includegraphics[width=1.0\linewidth]{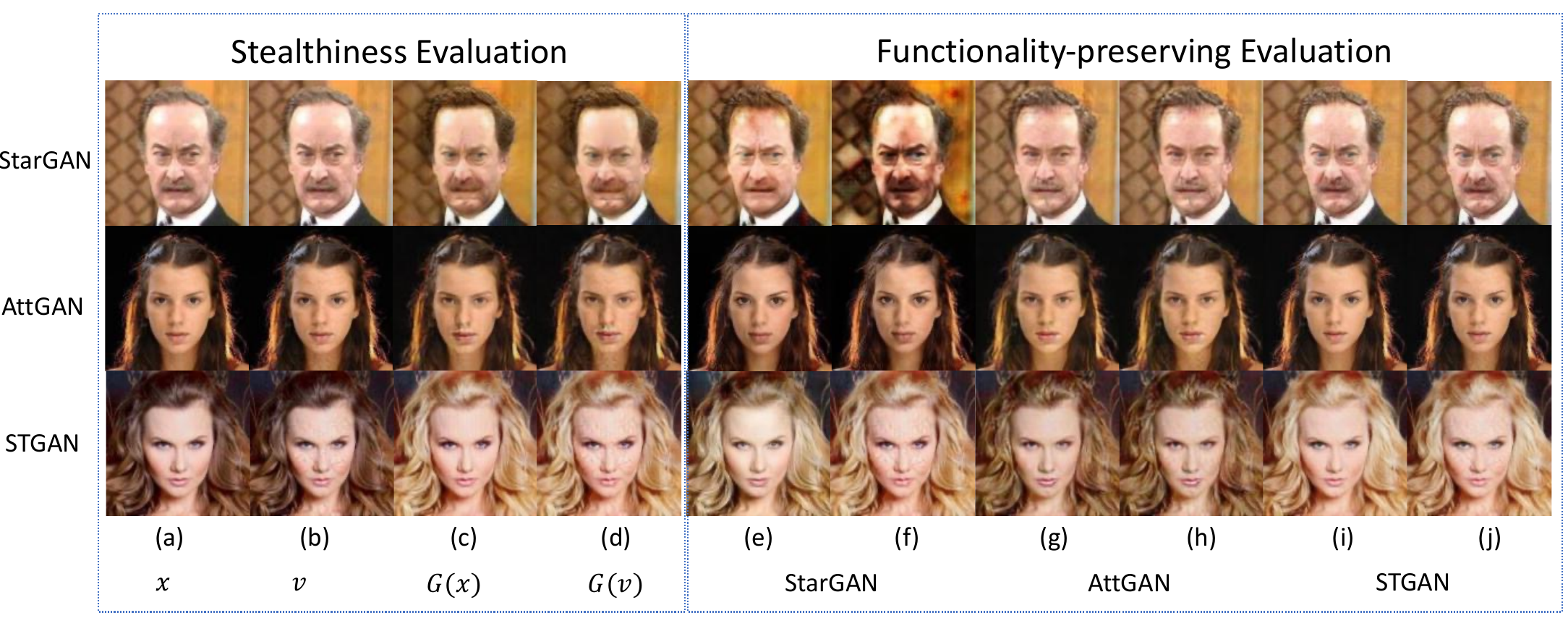}
\caption{Manipulated images. (e) edits attribute on $x$ with StarGAN, (f) edits attribute on $v$ with StarGAN, (g) edits attribute on $x$ with AttGAN, (h) edits attribute on $v$ with AttGAN, (i) edits attribute on $x$ with STGAN, (j) edits attribute on $v$ with STGAN.}
\label{fig:unique}
\vspace{-10pt}
\end{figure}

In Fig.~\ref{fig:unique}, we show our \texttt{CFP-iBDv2} verification samples' outputs for different GANs. The columns from (e) to (j) indicate the output images from different models manipulated on both clean samples and verification samples. Verification samples do not decrease other GANs outputs' quality in most cases. The outputs of verification samples look similar to outputs of clean samples. It means our \texttt{CFP-iBDv2} has good functionality-preserving property.

\begin{figure}[h]
\centering
\includegraphics[width=1.0\linewidth]{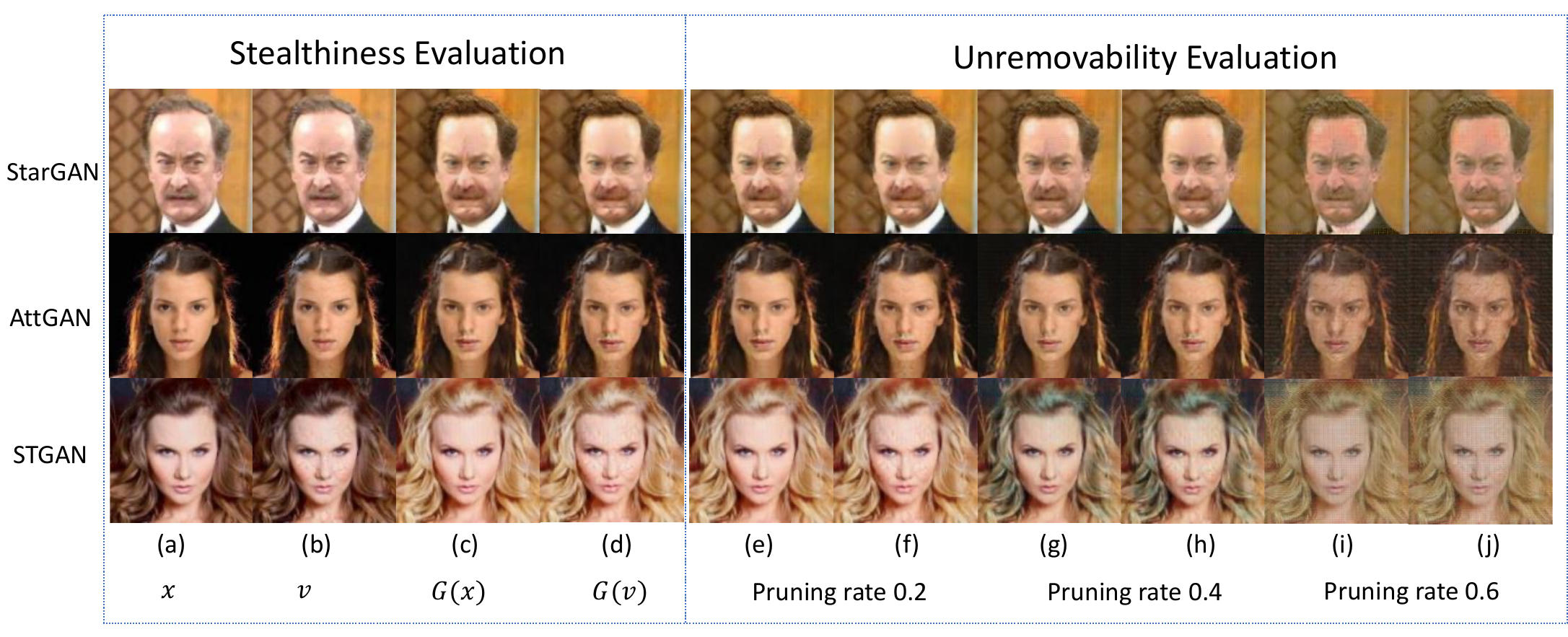}
\caption{Manipulated images. (e) edits attributes on $x$ with pruning rate $0.2$, (f) edits attributes on $v$ with pruning rate $0.2$, (g) edits attributes on $x$ with pruning rate $0.4$, (h) edits attributes on $v$ with pruning rate $0.4$, (i) edits attributes on $x$ with pruning rate $0.6$, (j) edits attributes on $v$ with pruning rate $0.6$.}
\label{fig:robustpruning}
\vspace{-15pt}
\end{figure}

\subsection {Verification Samples suffer Degradation}
\label{appendix:pruning}
In evaluating the unremovability of our proposed method, we mainly consider the two different degradation, model transformations and common image transformations. Here, we will give more results under different degrees of degradation.

\begin{figure}[h]
\vspace{-10pt}
\centering
\includegraphics[width=0.7\linewidth]{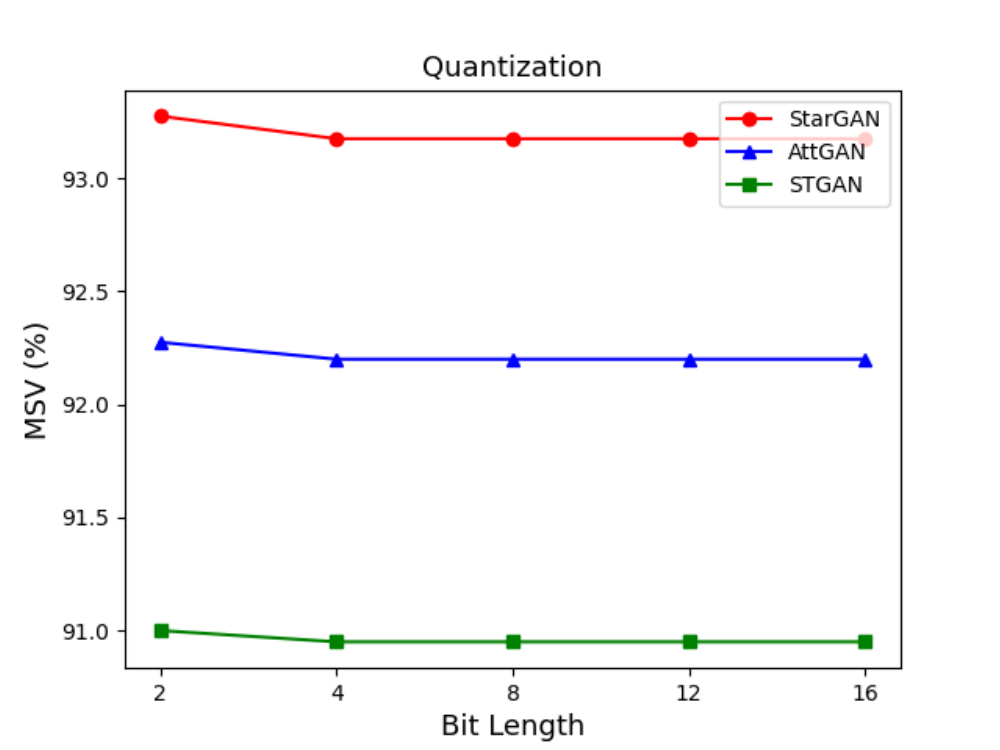}
\caption{MSV (\%) of \texttt{CFP-iBDv2} under different quantization scales. Bit Length stands for the truncation length for model parameters.}
\label{fig:quanti_rusult}
\vspace{-5pt}
\end{figure}

\begin{figure}[h]
\centering
\includegraphics[width=0.7\linewidth]{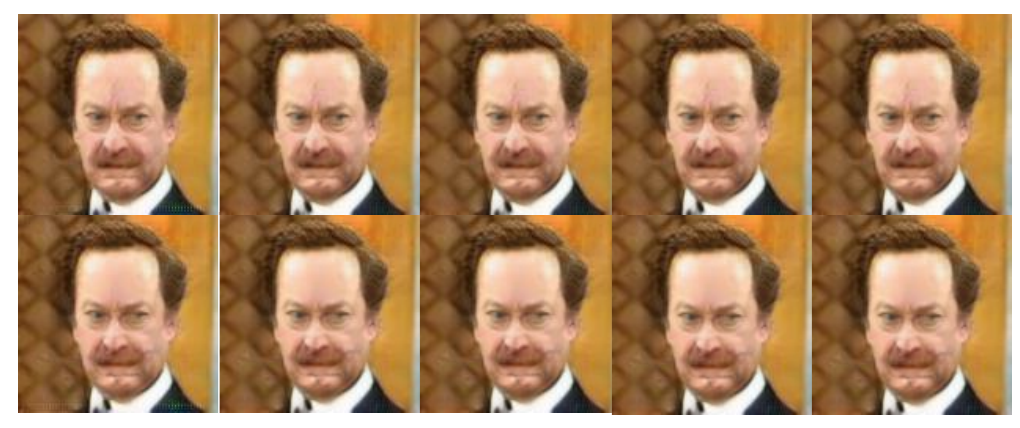}
\caption{Outputs before and after model quantization. The first row is the outputs of the clean image and the verification image, respectively. The second row is the corresponding outputs after model quantization. Each column responds to the output of one Bit Length quantization model in Fig.~\ref{fig:quanti_rusult}.}
\label{fig:quanti}
\vspace{-15pt}
\end{figure}

\begin{figure}[h]
\vspace{-5pt}
\centering
\includegraphics[width=0.7\linewidth]{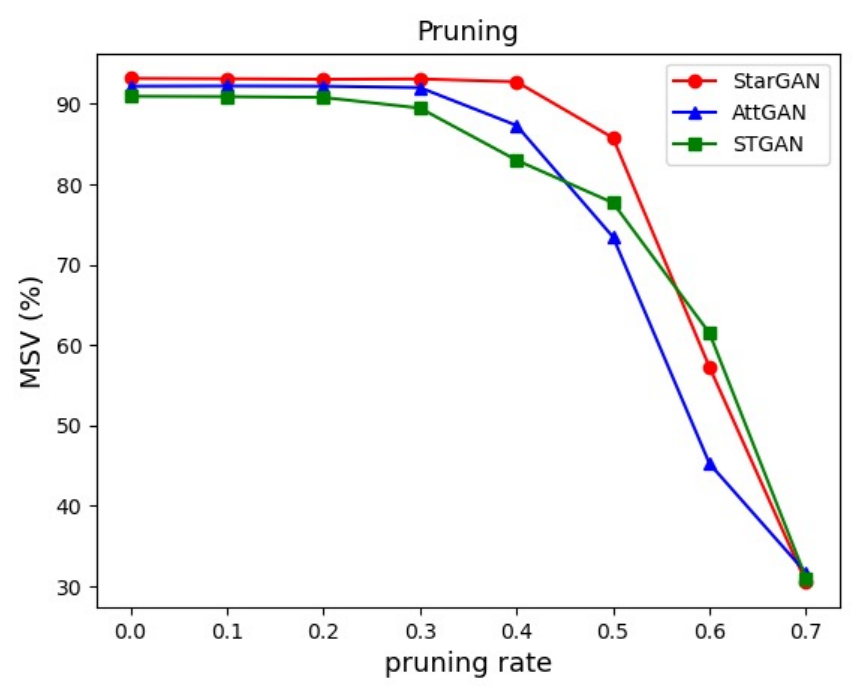}
\caption{MSV (\%) of \texttt{CFP-iBDv2} under pruning.}
\label{fig:pruing}
\vspace{-20pt}
\end{figure}

\noindent\textbf{Model Transformations.} In evaluating the unremovability against model compression (i.e., pruning), we explore the effectiveness of our method when the GAN model is compressed in various levels. In Fig.~\ref{fig:robustpruning}, the column (e) to (h) indicate the manipulated images with compressed models when the pruning rate is not larger than $0.4$. We can find that the GAN's outputs maintain a high-quality visualization, thus the pruning rate no more than $0.4$ is an appropriate setting in our experiment. Furthermore, if the pruning rate is higher, when it is 0.6, the outputs is not satisfying for a user. We further compare more experimental results under various pruning rates showing in the Fig.~\ref{fig:pruing}. When the pruning rate is smaller than 0.5, the MSV (\%) is high enough to pass the verification (the threshold is 0.8). With the pruning rate increasing, the MSV (\%) will drop slowly at first and decrease significantly after the pruning rate is higher than 0.5. Because the outputs' quality is not good enough for the backdoor classifier to recognize the triggers. We apply model quantization on GANs based on model parameter trunction, which means we keep model paramters with a specific length. After different scales' quantization, our method can successfully verify the fingerprinted GAN, which can be found in Fig.~\ref{fig:quanti_rusult}. The visulization results in Fig.~\ref{fig:quanti} indicates that under model quantization, GANs can generate high quality results.

\begin{figure}[h]
\centering
\includegraphics[width=1.0\linewidth]{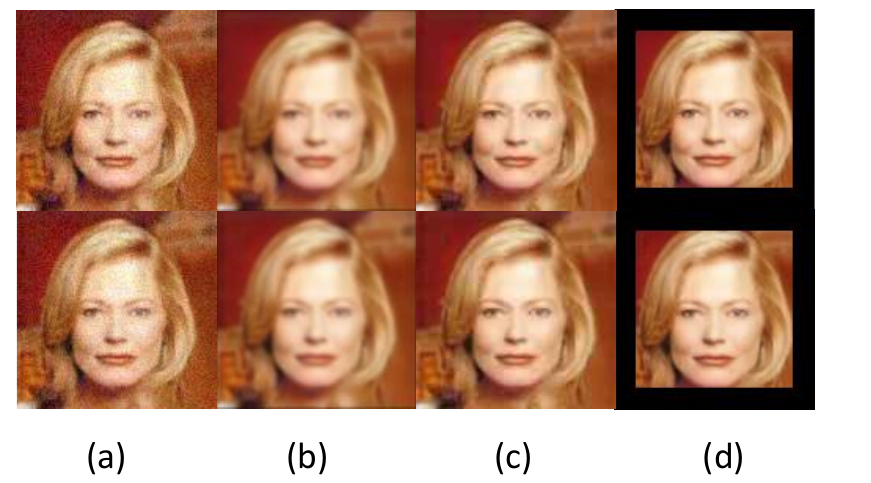}
\caption{Visualization of GAN outputs corrupted by four image transformations. The first row is GAN outputs from clean images. The second row is GAN outputs from verification samples. (a) adding Gaussian Noise, (b) blurring, (c) JPEG compression, (d) centering cropping.}
\label{fig:robustchange}
\vspace{-5pt}
\end{figure}

\begin{figure}[h]
\centering
\includegraphics[width=1.0\linewidth]{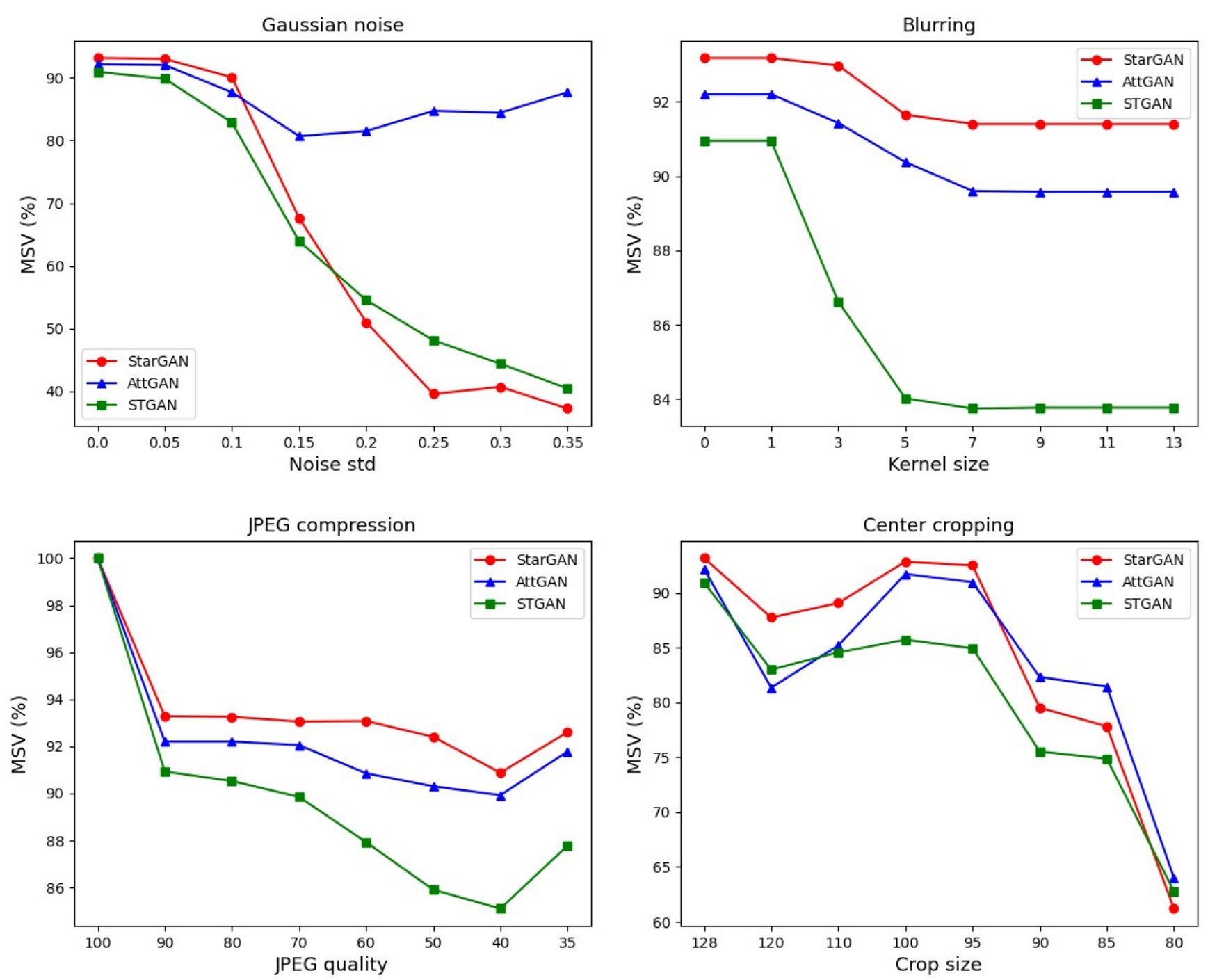}
\caption{MSV (\%) of \texttt{CFP-iBDv2} under different settings of transformations, applied on the outputs of GANs.}
\label{fig:odt}
\vspace{-10pt}
\end{figure}

\noindent\textbf{Image Transformations.} The GAN's outputs will always be corrupted by various image transformations when spreading in the social models. Fig.~\ref{fig:robustchange} presents the visualization of GAN's outputs by employing four different types of common image transformations, including adding \textit{Gaussian noises}, \textit{blurring}, \textit{JPEG compression}, and \textit{centering cropping}. Here, the parameters of these transformations are described in Section~\ref{sec:robust-analysis}. In Fig.~\ref{fig:odt}, we show the MSV (\%) under different transformation magnitudes, which transformation applies on the outputs of the GAN. Clearly, our \texttt{CFP-iBDv2} is robust under blurring and compression. These two transformations have trivial influence during the verification process. As for adding Gaussian noise, \texttt{CFP-iBDv2} is robust on the AttGAN, and when the noise std is higher than 0.1, the verification process will fail on the StarGAN and STGAN. Center cropping can significantly decline the completeness of backdoor triggers, resulting in the verification failure. Our \texttt{CFP-iBDv2} can still work when the cropping size is bigger than 90, which is an excellent result.

\begin{figure}[h]
\centering
\includegraphics[width=1.0\linewidth]{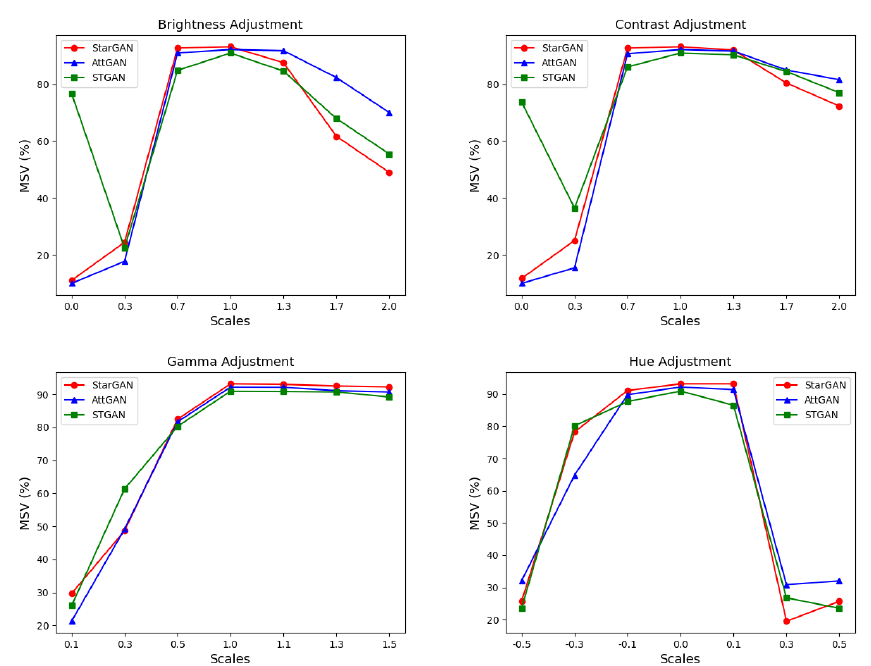}
\caption{MSV (\%) of \texttt{CFP-iBDv2} under different settings of transformations, applied on the outputs of GANs.}
\label{fig:odt_others}
\vspace{-15pt}
\end{figure}

\begin{figure}[h]
\centering
\includegraphics[width=1.0\linewidth]{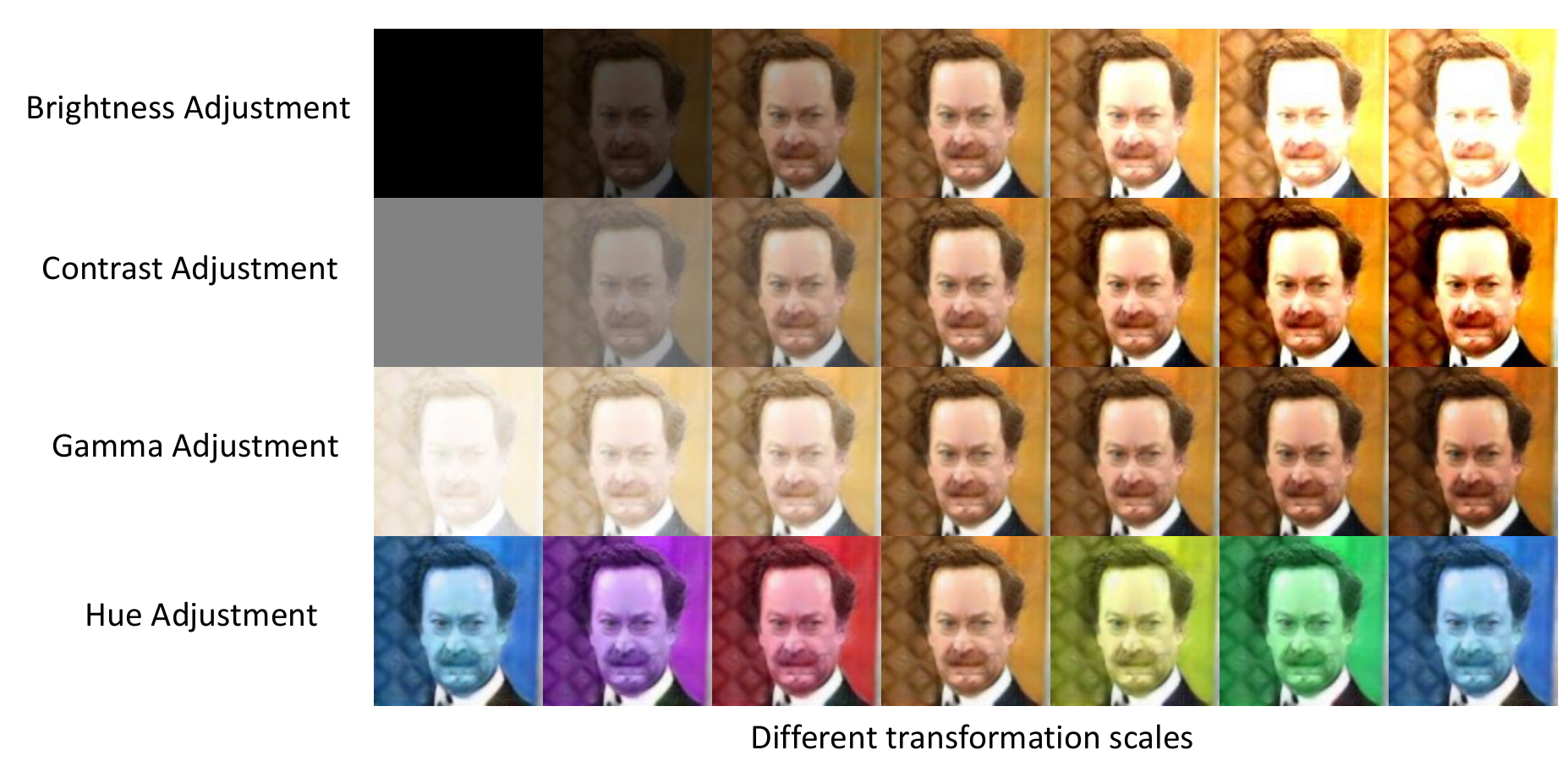}
\caption{Results of outputs after applying image transformations on the outputs. The transformation scales are consistent with the values on the x-axis in Fig.~\ref{fig:odt_others}.}
\label{fig:odt_others_vis}
\vspace{-10pt}
\end{figure}

Additionaly, we explore unseen image transformations' effects on our verification classifier. In Fig.~\ref{fig:odt_others}, we compare four unseen image transformations, i.e., brightness adjustment, contrast adjustment, gamma adjustment and hue adjustment. For each adjustment, we consider different transformation intensities, and the outputs after each can be found in Fig.~\ref{fig:odt_others_vis}. The results confirm that our method can defend against these unseen transformations even we do not use them to train our verification classifier.

\begin{figure}[h]
\centering
\includegraphics[width=1.0\linewidth]{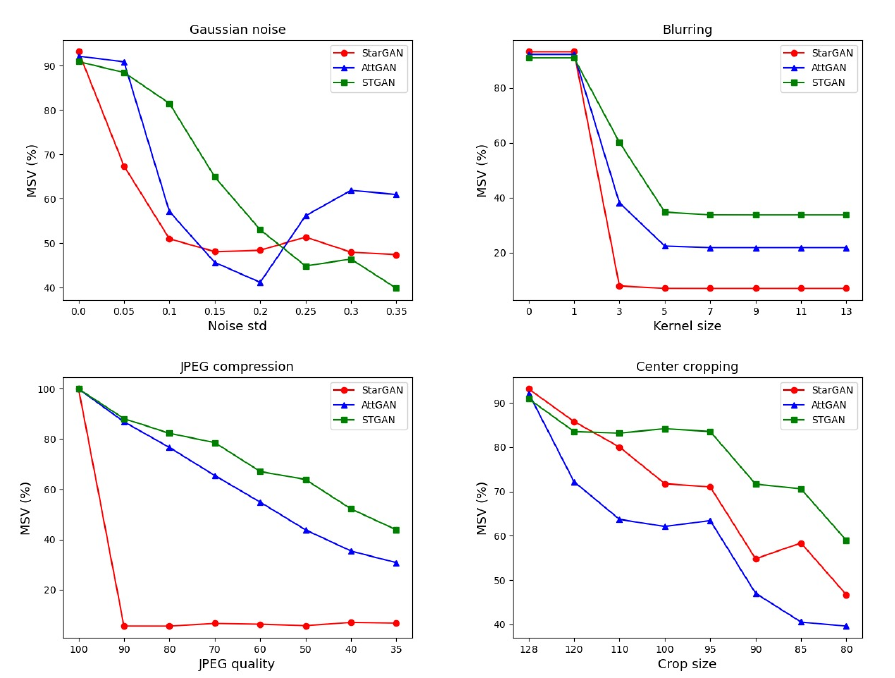}
\caption{MSV (\%) of \texttt{CFP-iBDv2} under different settings of transformations, applied on the inputs of the GAN.}
\label{fig:idt}
\vspace{-15pt}
\end{figure}

\begin{figure}[h]
\centering
\includegraphics[width=1.0\linewidth]{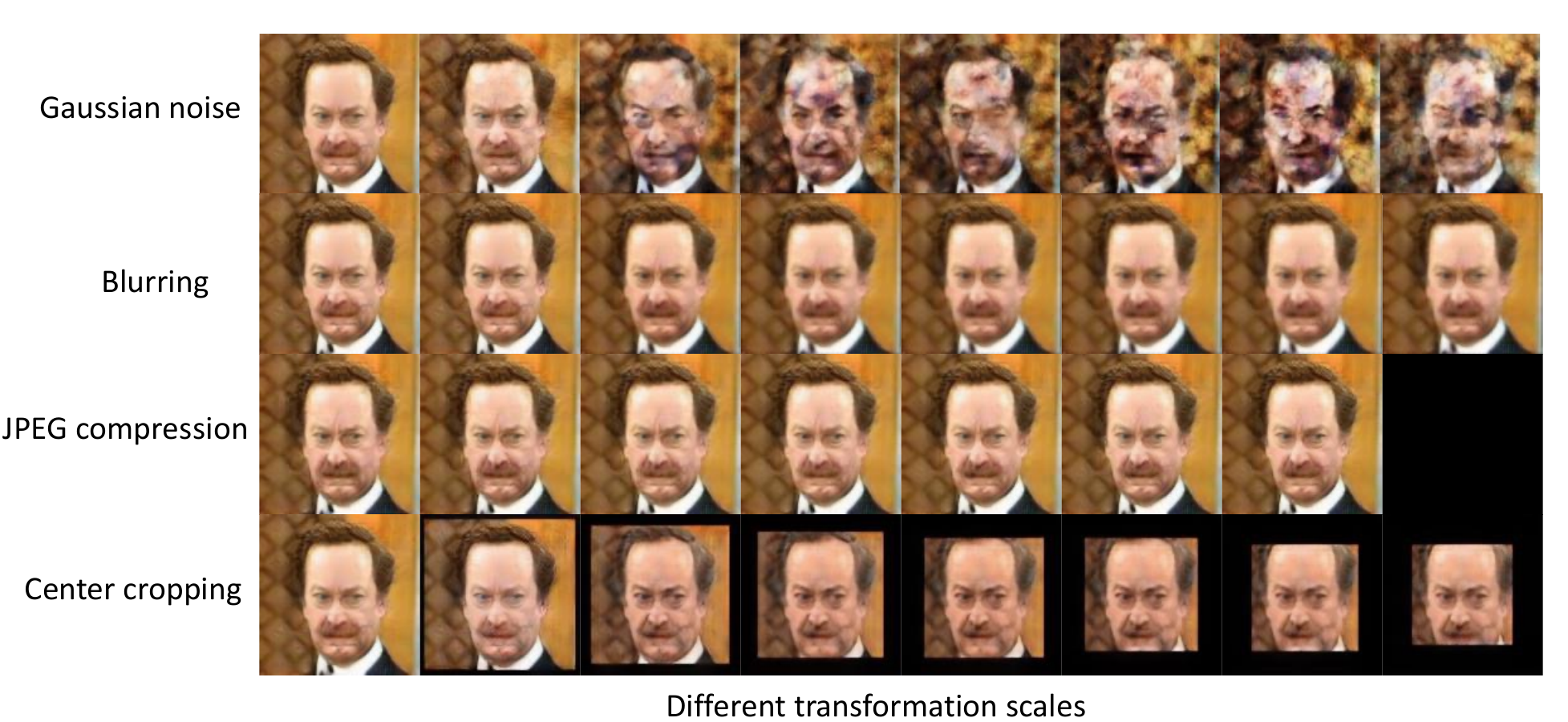}
\caption{Results of outputs after applying image transformations on the inputs. The transformation scales are consistent with the values on the x-axis in Fig.~\ref{fig:idt}.}
\label{fig:idt_vis}
\vspace{-20pt}
\end{figure}

Furthermore, in Fig.~\ref{fig:idt}, we show the MSV under different transformation magnitudes of \texttt{CFP-iBDv2}, in which transformation applies on the inputs of GANs. The MSV in these figures is significantly low, which is because when we add image transformations on the inputs, the outputs of the GAN lost most of details which contain the fingerprint information, which can be found in Fig.~\ref{fig:idt_vis}, especially under Gaussian noise and compression, which introduce non-trivial noise to replace our backdoor perturbation. We believe that this type of image transformations will not be used in practice as a defense.

After the comprehensive experiments on model pruning and image transformations, our \texttt{CFP-iBDv2} shows impressive \textit{functionality-preserving}, \textit{unremovability}, and \textit{stealthiness}. It can defend against gentle and medium image modification and model compression. More than that, its outputs are visually indistinguishable for humans.

\end{document}